\documentclass[aps,prb,twocolumn,superscriptaddress,showpacs,floatfix]{revtex4-1}
\usepackage{graphicx}
\usepackage{epsfig}
\usepackage{amsmath}
\usepackage{amssymb}
\usepackage{subfigure}

\begin{document}

\frenchspacing

\title{Simulation study of pressure and temperature dependence of the negative thermal expansion in Zn(CN)$_2$}

\author{Hong Fang}
\affiliation{Department of Earth Sciences, University of Cambridge, Downing Street, Cambridge CB2 3EQ, U.K.}

\author{Martin T. Dove}
\email{martin.dove@qmul.ac.uk}
\affiliation{Department of Earth Sciences, University of Cambridge, Downing Street, Cambridge CB2 3EQ, U.K.}
\affiliation{Centre for Condensed Matter and Materials Physics, School of Physics and Astronomy, Queen Mary University of London, Mile End Road, London E1 4NS, U.K.}

\author{Leila H. N. Rimmer}
\affiliation{Department of Earth Sciences, University of Cambridge, Downing Street, Cambridge CB2 3EQ, U.K.}

\author{Alston J. Misquitta}
\affiliation{Centre for Condensed Matter and Materials Physics, School of Physics and Astronomy, Queen Mary University of London, Mile End Road, London E1 4NS, U.K.}

\date{\today}
\begin{abstract}
Pressure and temperature dependence of the negative thermal expansion in Zn(CN)$_2$ is fully investigated using molecular dynamics simulations with a built potential model. The advantage of this study allows us to reproduce the exotic behaviours of the material, including the negative thermal expansion (NTE), the reduction of NTE with elevated temperature, the pressure enhancement of NTE and the pressure-induced softening. Results of the study provide us detailed data to link the properties in the energy space and the real space, giving us insights to understand the properties and the connections between them.
\end{abstract}


\maketitle

\section{INTRODUCTION}

Negative thermal expansion (NTE) is a rare and counter-intuitive phenomenon found primarily in low density materials with crystal structures that are networks of linked coordination polyhedra. The study of these materials is not only of fundamental scientific importance, but also has many technological applications such as aerospace technologies~\cite{Imanaka 2000}, optics~\cite{Clegg 2002}, and electronics~\cite{Closmann 1998}. While much effort has been put into finding new materials and investigating the origin of NTE in them, much less attention has been paid to the change in NTE behaviour subject to heating and stress, which holds great importance for the possible applications of the material. For example, due to stresses and heating, problems such as phase transitions of the NTE filler and thermal expansion misfit between the NTE filler and matrix are always encountered in designed composites with tailored thermal expansion \cite{Hermann 1999,Takenaka 2012}.

In this paper, we conduct a simulation study of Zn(CN)$_2$ focusing on the pressure and temperature effects on its negative thermal expansion. We chose this material for several reasons. Firstly, Zn(CN)$_2$ is a well-known representative NTE material \cite{Goodwin 2005}. It has a framework structure consisting of tetrahedral groups of atoms linked by diatomic rods of C--N and has exceptionally large isotropic NTE of $\alpha_\mathrm{linear}=-16.9$ MK$^{-1}$ (twice as large as that of ZrW$_2$O$_8$ \cite{Mary 1996}). Secondly, the material shows a variety of exotic properties in experiments \cite{Chapman 2005,Chapman 2007}, including reduction of its NTE on heating, pressure-enhanced thermal contraction, and pressure-induced softening, none of which are fully understood. Thirdly, with previous DFT calculations of Zn(CN)$_2$ \cite{Zwanziger 2007,Ravindran 2007,Ding 2008,Mittal 2011} explaining the origin of NTE of the material in terms of Gr\"{u}neisen theory, it would be useful to draw a clear link between the values of the Gr\"{u}neisen parameters in energy space and the structural vibrations in real space with full anharmonicity (which should be important in such an NTE system) based on theoretically reproducing the aforementioned exotic properties.

Here, we have built a Zn(CN)$_2$ potential model based on first-principles calculations. Lattice-dynamic calculations and large-scale molecular dynamics (MD) simulations were carried out for the material using this model which was justified by comparing against the available experimental data. The results in both energy and real space provide us fundamental clues to understand the NTE as well as the related exotic behaviours of Zn(CN)$_2$.

\begin{figure}[t]
\centering
\subfigure{\includegraphics[width=8.0cm]{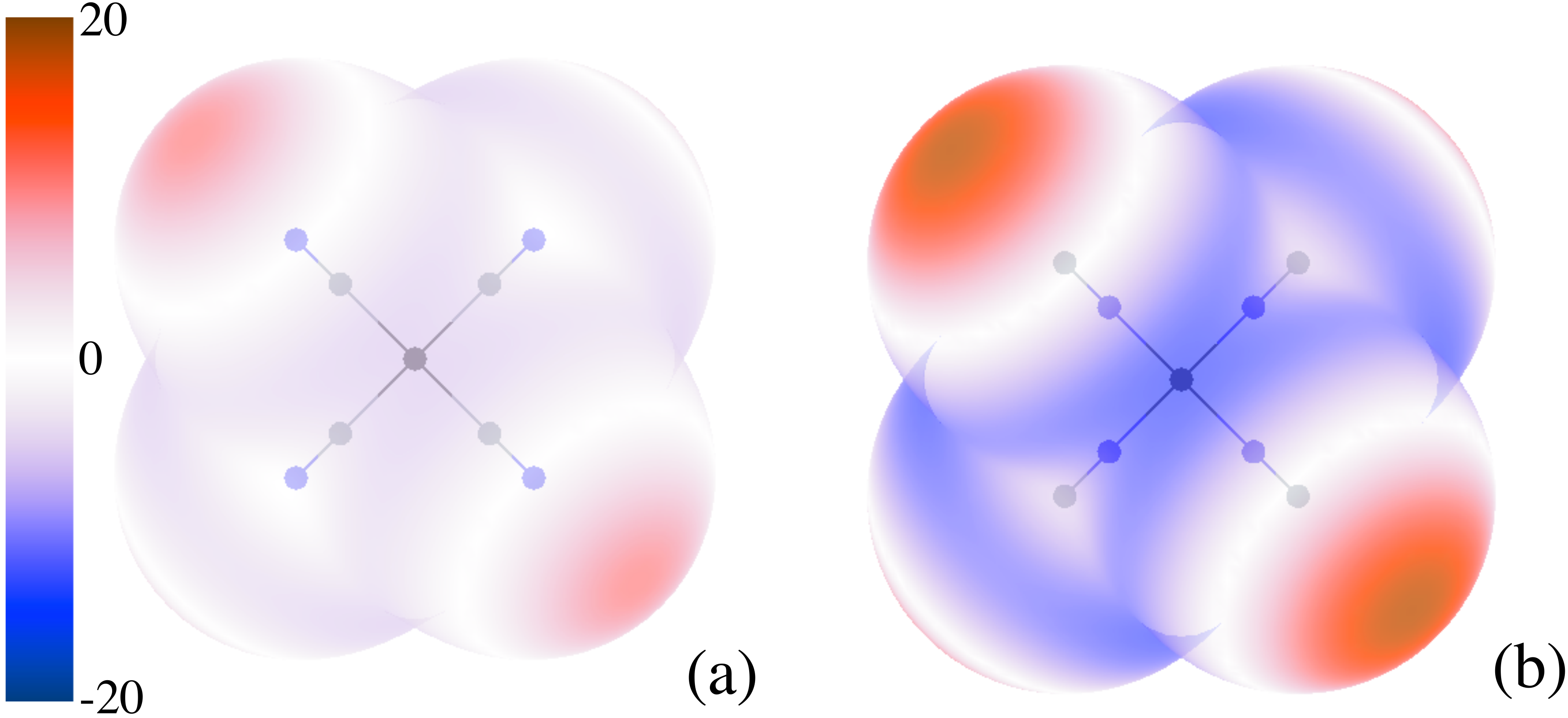}}
\label{fig1}
\caption{\label{fig1} Plot showing the difference between the electrostatic potentials arising from rank 4 DMA and rank 0 Mulfit  \cite{Ferenczy 1991,Ferenczy 1998}, visualised on the van der Waals surface of (a) a Zn--C--N ordered cluster and (b) a Zn--N--C ordered cluster. The colour bar shows values of the electrostatic potential difference in kJ/mol. The average percentage error in the electrostatic potential made by the point charge model on this surface is 1\%.}
\end{figure}

\section{Building the model}

\begin{table*}[t]
\caption{The potential model. E is energy, $r$ is inter-atomic distance, $r_0$ is equilibrium inter atomic distance, $\theta$ and $\varphi$ are bond angles, and $\theta_0$ is equilibrium bond angle. Energies are in eV, distances are in \AA~and bond angles are in degrees. The prefactor of the harmonic bond-bending term is in eV/rad. }
\centering
\begin{tabular}{c c c c}
\hline\hline
Potential & Form of the Potential & Type of Bond & Values of Parameters \\[0.5ex]
\hline

& & Zn--C & $D = 0.2432$, $\alpha = 2.917$, $r_0 = 2.109$  \\[-1ex]
\raisebox{1.5ex}{Morse potential} & \raisebox{1.5ex}{$E_M = D\{\left[1-\exp\left(-\alpha(r-r_0)\right)\right]^{2}-1\}$}
& Zn--N & $D = 0.1795$, $\alpha = 2.701$, $r_0 = 2.157$  \\[1ex]

& & C--Zn--C & $K = 1.5157$, $\theta_0 = 109.47$ \\[-1ex]
Harmonic three-body potential & $E_H = (1/2)K(\theta-\theta_0)^{2}$ & C--Zn--N & $K = 1.2695$, $\theta_0 = 109.47$ \\[-1ex]
& & N--Zn--N & $K = 1.0233$, $\theta_0 = 109.47$ \\[1ex]

& & Zn--C--N & $K = 1.1968$ \\[-1ex]
\raisebox{1.5ex}{Linear three-body potential} & \raisebox{1.5ex}{$E_L = K(1-\cos\varphi)$}
& Zn--N--C & $K = 0.6359$ \\[1ex]

& & C--C & $A = 2806.9$, $\rho = 0.2667$, $C = 17.67$ \\[-1ex]
Buckingham potential & $E_B = A\exp(-r/\rho)-(C/r^{6})$ & C--N & $A = 2365.0$, $\rho = 0.2825$, $C = 20.88$ \\[-1ex]
& & N--N & $A = 1992.7$, $\rho = 0.2874$, $C = 24.67$ \\[1ex]

\hline
\end{tabular}
\label{table1}
\end{table*}

We started with total-energy calculations for a [Zn(CN)$_4$]$^{2-}$ cluster. The $-2$ charge comes from the fact that every zinc is shared by four neighbour atoms (C or N), and that the bond should be highly ionic according to an initial judgment of zinc having much smaller electronegativity than carbon and nitrogen. Four CN$^{-}$ molecular ligands around each Zn$^{2+}$ would give us a $-2$ charge on the cluster. Geometry optimization of the cluster resulted in a perfect tetrahedral conformation.

Total energies for different configurations of the cluster (with bond stretching and angular distortions) were computed using DFT in GAMESS(US) with the PBE0 functional \cite{Schmidt 1993,Gordon 2005}. Correlation-consistent basis sets up to aug-cc-pVQZ were tested, and an aug-cc-pVTZ basis set was found to have sufficient accuracy and no significant basis set superposition errors (BSSE) at different configurations.

Various interatomic potential forms were then fitted to the calculated energy curves to obtain the initial potential parameters. The short-range interactions are the Morse potential for Zn--C/N, a harmonic three-body-bond-bending term for C/N--Zn--N/C and a linear-three-body term for the angular distortion of Zn--C/N--N/C. The long-range van der Waals interactions are described by a Buckingham potential with parameters from Williams \cite{Williams 2001}.

Two types of clusters with Zn--C--N order and Zn--N--C order are used. The multipoles on each cluster were calculated by distributed multipole analysis (DMA) \cite{Stone 2005} using Cam\uppercase{CASP} \cite{Misquitta 2012}. The effective point charges on the atoms were then obtained by fitting to the electrostatic potential from the rank 4 (hexadecapole) distributed multipoles using the MULFIT program \cite{Ferenczy 1991,Ferenczy 1998}. Fig.~\ref{fig1} shows the difference in the electrostatic potential between the point-charge model and the DMA result. The root mean square of the difference is less than $4$ kJ/mol and $8$ kJ/mol for clusters of Zn--C--N and Zn--N--C, respectively, corresponding to about 1\% relative difference in the electrostatic potential around the clusters. The averaged effective point charges on each atom are (in electron units) $+1.14$ for Zn, $-0.21$ for C and $-0.36$ for N.

The initial potential model was then refined by refitting to DFT energy surfaces of various cluster configurations. The point charges were fixed during this process.

This gave us final potential parameters that implicitly incorporate the effects of higher-ranking multipole moments and atomic polarization. Due to the high strength of the C--N bond \cite{Ludi 1973, Sharpe 1976, Dunbar 1997, Williams 1997, Verdaguer 2004}, this group was treated as a rigid rod in all cases. The potentials with their parameters are listed in Table~\ref{table1}.

\section{Computational Methods}

Harmonic lattice dynamics (HLD) and quasi-harmonic lattice dynamics (QHLD) calculations were carried out using GULP \cite{Gale 1997}.

Molecular dynamics (MD) simulations were performed using DL$\_$POLY \cite{Todorov 2006} for a $10\times10\times10$ supercell containing $10000$ atoms with periodic-boundary conditions. A constant stress constant temperature (N$\sigma$T) ensemble with a Nos\'{e}-Hoover thermostat \cite{Hoover 1985} was used. The long-range Coulomb interactions were calculated using the Ewald method with precision of 10$^{-6}$. The equations of motion were integrated using the leapfrog algorithm with a time step of $0.001$ ps. A total of $20000$ time steps were used to achieve equilibration. At different temperatures and pressures, snapshots of atomic trajectories after equilibration were recorded every $0.02$ ps up to a total of $50$ ps for the follow-up analysis. Both ordered model with $P\overline 43m$ symmetry and disordered model with $Pn\overline3m$ were used. The latter was constructed by randomly switching C and N atoms in the supercell.

\section{An Initial Test of the Model}

We compared some basic quantities obtained from the model against experiment. The optimized structure gave the cell parameter as $a=5.9176$ {\AA}, and the Zn--C/N bond length as $r=1.9782$ {\AA} compared to $a=5.9227(1)$ {\AA} and $r=1.9697(3)$ {\AA} from an experiment at $14$ K \cite{Williams 1997}. The model is found to be stable in lattice-dynamic calculations, and the calculated phonons are in good agreement with the spectroscopy data \cite{Ravindran 2007}, as shown in Table~\ref{freqtable}. The dispersion curves shown in Fig.~\ref{fig:freqsoft} generally agree well with previous DFT calculations \cite{Zwanziger 2007,Mittal 2011}.

\begin{table}[t]
\caption{Calculated phonon frequencies at the $\Gamma$ point for both CN ordered and CN disordered (virtual crystal) models. Experimental Infra-red and Raman~\cite{Ravindran 2007} data and DFT calculated results~\cite{Mittal 2011} are provided for comparison. Frequencies are in cm$^{-1}$.}
\centering
\begin{tabular}{p{1.8cm} p{1.2cm} p{1.2cm} p{1.2cm} p{1.2cm} p{1.2cm}}
  \hline\hline
  Mode &T$_{1u}$ & T$_{2g}$ & E$_g$ & T$_{2g}$ & T$_{1u}$ \\[0.5ex]
  \hline
  Ordered Model  & 173  &  186  &  328  &  337  &  463  \\ [0.5ex]
  Disordered Model  & 210 & 220 & 314 & 349 & 490 \\ [0.5ex]
  Experiment &  178  &  216  &  334  &  339  &  461  \\ [0.5ex]
  DFT~\cite{Mittal 2011} &  178  &  204  &  330  &  336  &  476  \\ [0.5ex]
  \hline
\end{tabular}
\label{freqtable}
\end{table}

It is worth noting the close similarity between the dispersion relations of wave vectors along $\Gamma$--X--R and R--M--$\Gamma$, as well as the near-mirror symmetry of the dispersion curves along $\Gamma$--R (the middle line of $\Gamma$--R is the mirror line). This makes sense given the existence of the two interpenetrating cristoablite-like networks in the material. For the lowest-frequency mode at wave vector R, the two networks would translate like acoustic modes but out of phase with each other, leading to a dispersion relation that has the appearance of an acoustic mode and a positive Gr\"{u}neisen parameter but with a non-zero band gap. The frequency of this mode is 0.16 THz, lower than the values of 0.69 THz and 0.48 THz from the DFT calculations in Ref.~\onlinecite{Zwanziger 2007} and Ref.~\onlinecite{Mittal 2011}, respectively, suggesting a softer long-range interaction between the two networks in our model.

\begin{figure}[t]
\centering
\subfigure{\includegraphics[width=8.0cm]{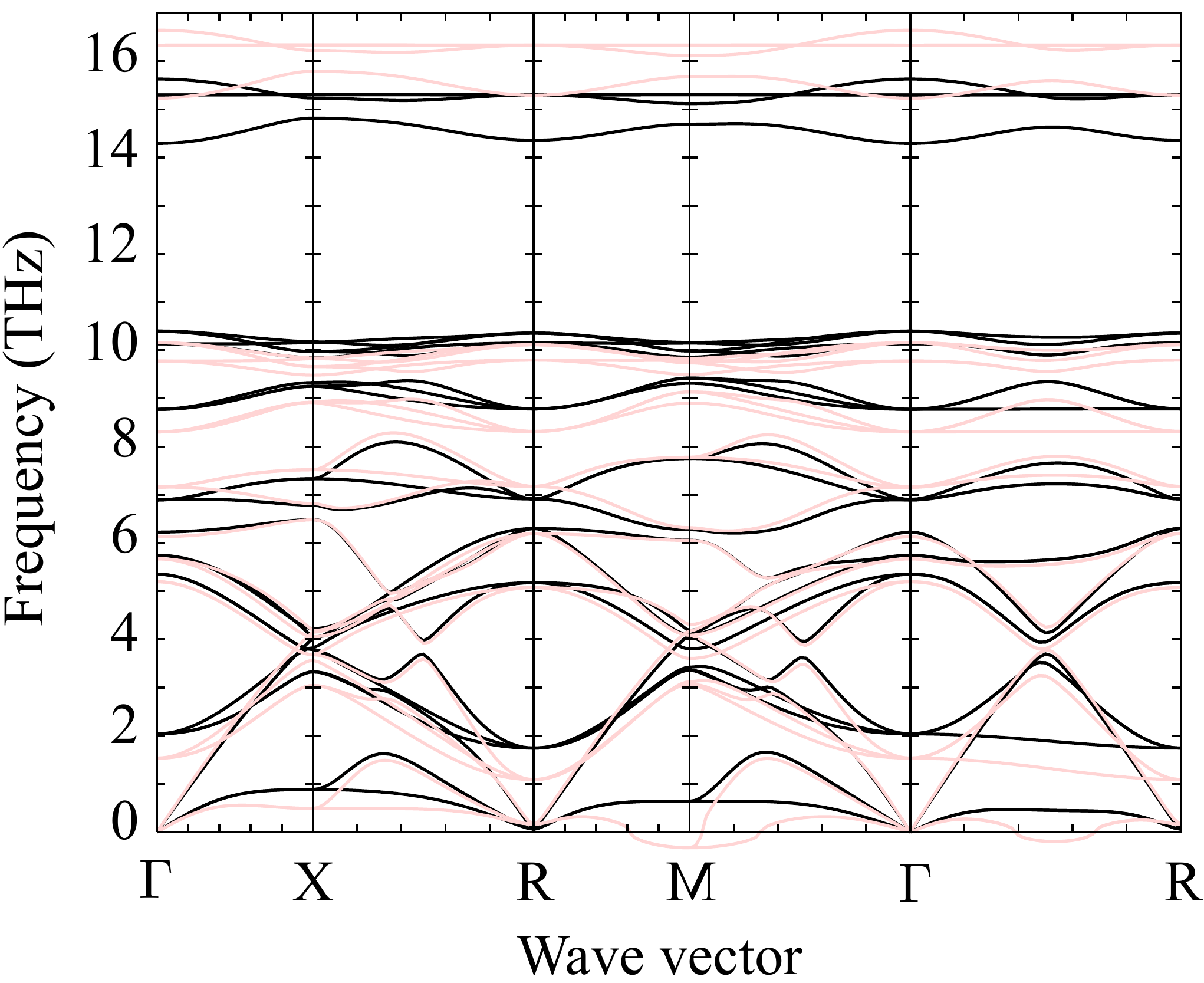}}
\label{fig:freqsoft}
\caption{\label{fig:freqsoft} Zn(CN)$_2$ dispersion curves calculated along high-symmetry directions in the Brillouin zone. Black curves are the phonons of the zero-pressure cell; light-red curves are the phonons of the cell at 1.0 GPa hydrostatic pressure. The acoustic modes at M (0.5, 0.5, 0.0) and at the midpoint of $\Gamma$--R (0.25, 0.25, 0.25) are the first to become unstable at high pressure.}
\end{figure}

\section{Thermodynamic properties}

\subsection{Pressure and temperature dependence of the NTE}

\begin{figure}[t]
\centering
\subfigure{\includegraphics[width=8.0cm]{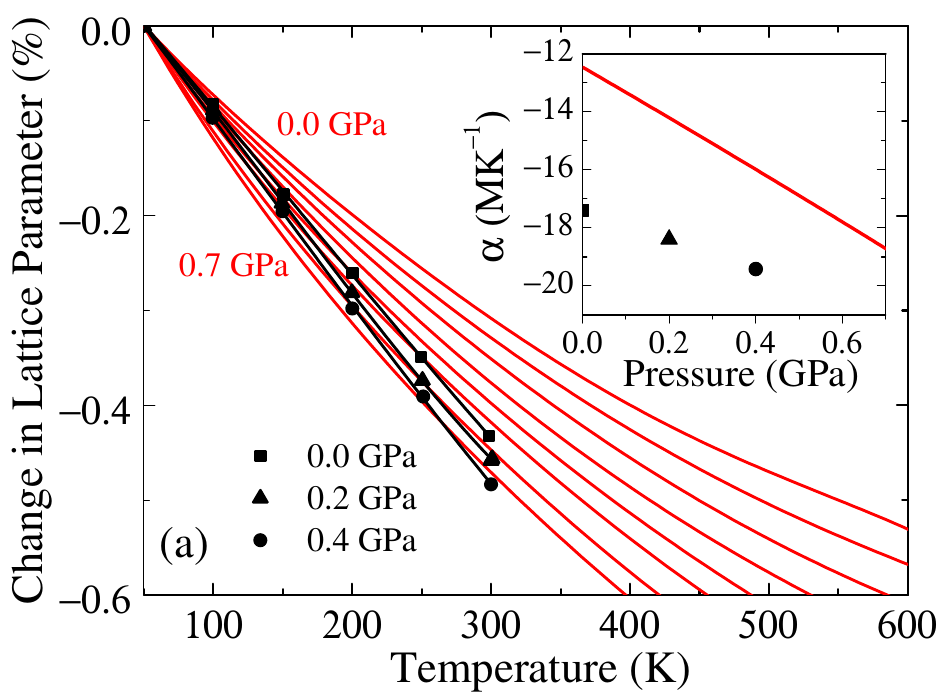}}
\label{fig3}
\caption{\label{fig3} Zn(CN)$_2$ isotherms calculated for pressures $0.0$--$0.7$ GPa in $0.1$ GPa increments. Red lines are calculated values; black markers are from neutron diffraction data\cite{Chapman 2007}. The inset shows the enhancement of NTE under pressure, observed in both calculations and in experiment.}
\end{figure}

\begin{figure}[t]
\centering
\subfigure{\includegraphics[width=8.0cm]{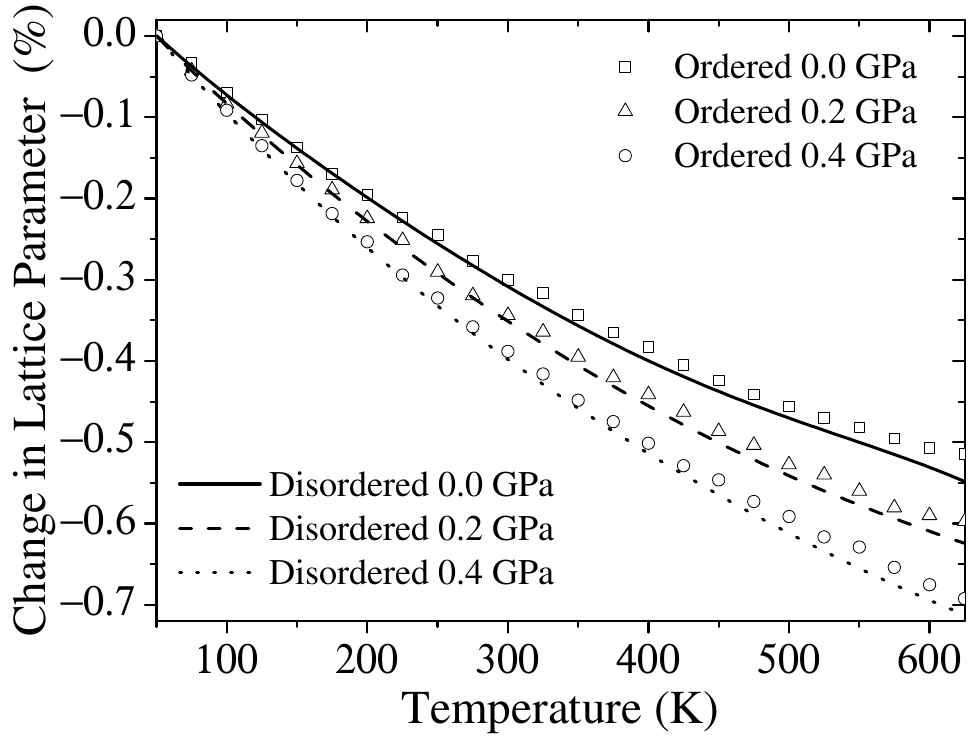}}
\label{fig3c}
\caption{\label{fig3c} NTE in Zn(CN)$_2$ calculated using MD. Symbols are data from the CN-disordered system, solid lines are data from the CN-ordered system. CN ordering thus has a negligible effect on the NTE behaviour.}
\end{figure}

The calculated NTE curves of Zn(CN)$_2$ under different pressures from $0.0$ to $0.7$ GPa with increments of $0.1$ GPa are shown in Fig.~\ref{fig3}. The pressure-enhanced $\alpha$ (averaged over $50$--$300$ K) at $0.0$, $0.2$ and $0.4$ GPa are $-12.62$, $-14.09$ and $-15.97$\,MK$^{-1}$, respectively, compared to the experimental values \cite{Chapman 2007} of $-17.40(18)$, $-18.39(27)$ and $-19.42(23)$\,MK$^{-1}$ at the corresponding pressures. The MD successfully captured the gradual reduction of NTE on heating which has been observed in X-ray scattering~\cite{Chapman 2005}. According to the MD, $\alpha_\textrm{V}$ is $-30.3$ MK$^{-1}$ at 300 K, much lower than the value at 25 K, $-47.3$~MK$^{-1}$.

Experiment \cite{Williams 1997} has confirmed that Zn(CN)$_2$ exists in a disordered form with $Pn\overline3m$ symmetry, i.e. the carbon and nitrogen atoms are randomly placed on their symmetric sites in the structure, compared to the ordered model with $P\overline 43m$ symmetry. NTE of the disordered model from the MD calculation is displayed as symbol plots in Fig.~\ref{fig3c} compared to that of the ordered model in solid lines. Clearly, there is no significant difference between the NTE of these two systems.

\subsection{Pressure and temperature dependence of the mechanical properties}

\begin{figure}[t]
\begin{center}
\includegraphics[width=8.2cm]{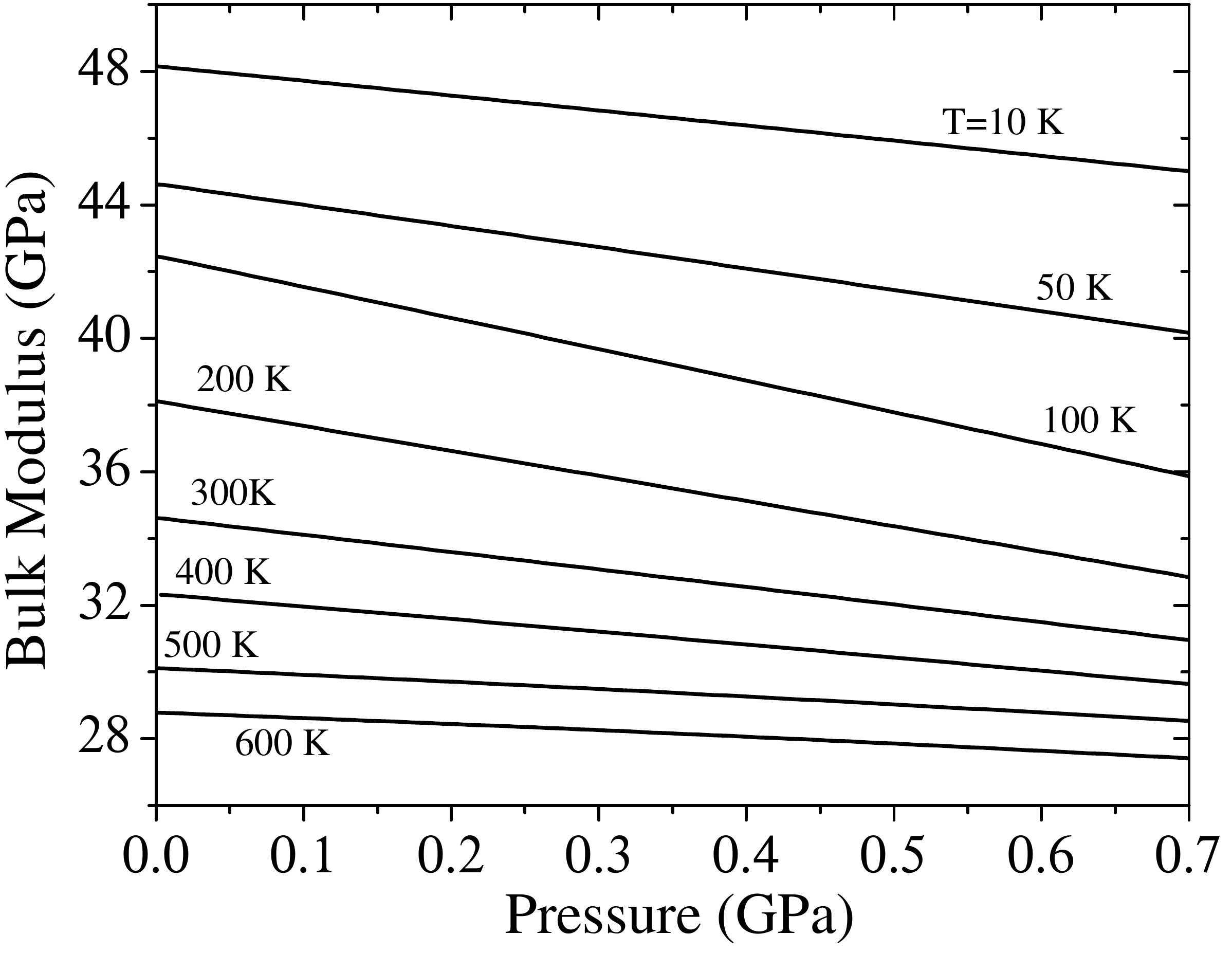}
\end{center}
\caption{\label{fig5} Pressure and temperature dependence of the bulk modulus of Zn(CN)$_2$. Results calculated using the $p$--$V$ data from MD simulations and standard thermodynamic relations. Pressure has far less influence on the bulk modulus than temperature.}
\end{figure}

The bulk modulus $B$ is directly related to the coefficient of thermal expansion $\alpha_\mathrm{v}$ according to the classical form

\begin{equation}\label{eq1}
\alpha_\mathrm{V} = \frac{{\overline \gamma c_v }}{{B}}
\end{equation}

\noindent where $c_v  = \sum\limits_s {\left( {\hbar \omega _s /V} \right)\left( {\partial n_s /\partial T} \right)}$ is the specific heat and the sum is over all modes with frequencies $\{\omega_s\}$. $n_s  = \left[ {\exp (\hbar \omega _s /k_B T) - 1} \right]^{ - 1}$ is the Bose-Einstein relation with $k_B$ the Boltzmann constant. The overall Gr\"{u}neisen parameter $\overline \gamma$ is calculated by summing over all mode Gr\"{u}neisen parameters weighed with their contribution to the specific heat.

At 300 K, the values of the bulk modulus $B_0$ and its first derivative with respect to pressure $B_0^{\prime}$ at zero pressure obtained by using the $3$rd-order Birch-Murnaghan equation of states to fit to the isotherms calculated from our MD are $B_0 = 34.47(31)$ GPa and $B_0^{\prime} = -4.2(5)$. Both are in good agreement with the experimental values of $34.19(21)$ GPa and $-6.0(7)$, respectively \cite{Chapman 2007}. However, the values of the bulk modulus obtained from the previous DFT calculations \cite{Zwanziger 2007,Mittal 2011} are much higher than the experiment.

We obtained the bulk modulus of Zn(CN)$_2$ at different temperatures and pressures by numerically computing derivatives of the $p$--$V$ data from the MD simulations. As shown in Fig.~\ref{fig5}, on cooling from 300 K to 50 K, $B$ increases by 26\% compared to the experiment of 15\% \cite{our_Exp}. Note that $B$ doesn't change much with pressure, but changes largely with temperature. Referring to Fig.~\ref{fig3}, we found that a $5$\% decrease in volume caused by heating at zero pressure corresponds to as much as $60$\% decrease in the bulk modulus, while the same amount of volume decrease caused by compression would only reduce the bulk modulus by less than $5$\%. This means that the bulk modulus of the material not only depends on the volume change per se, but also on the means of changing the volume --- by heating or compression. This breaks Birch's law of corresponding states~\cite{Birch 1961,Anderson 1965,Anderson 1967}. The same anomaly has been observed experimentally in ZrW$_2$O$_8$, where the bulk modulus increases by $40$\% on cooling from $300$ K to $0$ K \cite{Drymiotis 2004}.

\begin{figure}[t]
\begin{center}
\includegraphics[width=8.0cm]{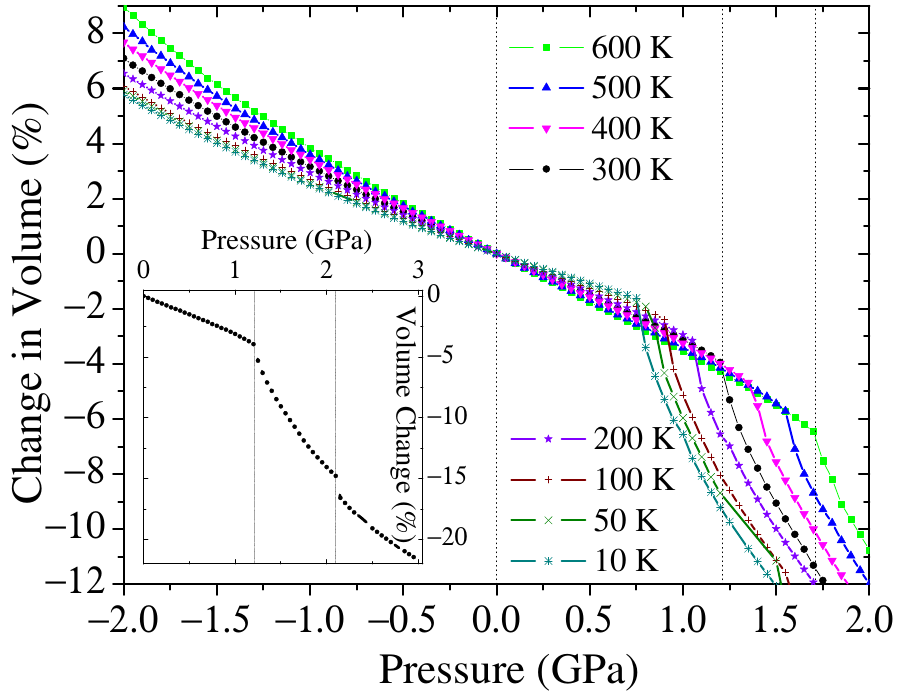}
\end{center}
\caption{\label{fig6} Pressure and temperature-dependence of Zn(CN)$_2$ volume, calculated using MD. The inset shows the change in volume with pressure up to $3.0$ GPa at $300$ K. A high-pressure phase settles down after the clear discontinuity at $2.1$ GPa.}
\end{figure}

According to the thermodynamic expressions of $\alpha_\mathrm{V}=\partial (\ln V) / \partial T$ and $B=\partial p/ \partial (\ln V)$ combined with Maxwell relation $\partial^2 V / (\partial T \partial p)_{p,T}= \partial^2 V / (\partial p \partial T)_{T,p}$, the pressure enhanced NTE of the material follows naturally from the relation

\begin{equation}\label{eq4}
\left( {\frac{{\partial \alpha_\mathrm{V} }}{{\partial p}}} \right)_T  = \frac{1}{{B^2 }}\left( {\frac{{\partial B}}{{\partial T}}} \right)_p.
\end{equation}

\noindent Since the temperature dependence of $B$ is negative as shown in Fig.~\ref{fig5}, $\alpha_\mathrm{V}$ would become more negative on compression. This is consistent with what seen in the positive expansion materials where $\alpha_\mathrm{V}>0$ so that $\alpha_\mathrm{V}$ would decrease on compression.

We found that the value of $B_0^{\prime}$ from a lattice-dynamic calculation using GULP \cite{Gale 1997} is $7.2$ compared to its negative value at 300 K. This suggests that all the mechanical contributions at $T=0$ are from the Zn--C/N bonds that become stiffened on compression. With elevated temperature, one can imagine that the Zn--C/N--N/C angle flexing starts to contribute to the change of volume on pressure. This mechanism costs much less energy than compressing the Zn--C/N bonds as suggested by values of the parameters in the Morse potential and the linear-three-body potential of the model (Table~\ref{table1}). As a result, the bulk modulus decreases on compression hence the negative $B_0^{\prime}$, i.e. pressure-induced softening of the material. At high temperature, $B_0^{\prime}$ is expected to become less negative on heating due to the rising energy cost of further increasing the Zn--C/N--N/C angle vibrational amplitude on compression. The temperature dependence of $B_0^{\prime}$ is specially discussed in our other work \cite{our_Exp}.

\section{Phase transitions}

In Fig.~\ref{fig6}, simulations were conducted for Zn(CN)$_2$ at different temperatures in a much broader range of pressure from $-2.0$ to $3.0$ GPa, and there is clearly a phase transition of the material caused by compression at each temperature. With elevated temperature, the phase-transition pressure indicated by the discontinuity in the volume change increases, implying that the phase transition may be triggered by some soft modes that can be stabilized on heating due to the anharmonic term in their frequencies. Indeed, as shown in Fig.~\ref{fig:freqsoft}, if we compare the dispersion curves calculated at a pressure beyond 1.0 GPa (in light red) to that at zero pressure (in black), we find softening of the acoustic modes around the zone boundaries, especially the modes at M and the mid-point of $\Gamma$--R $(0.25, 0.25, 0.25)$ which are the first ones to become unstable. The concurrent softening of the optic modes ($\sim 1.5$ THz) directly above these acoustic modes suggests a possible hybridization between the acoustic modes and the optic modes, resulting in a $k^2$ energy behaviour \cite{Martin 2000} at $\textbf{k}\neq0$. Same result was found for the disordered model.

The inset of Fig.~\ref{fig6} shows the change of volume with pressure up to $3.0$ GPa at $300$ K. The volume discontinuities at $1.2$ and $2.1$ GPa may suggest hysteresis in the phase transition. We found that the new high-pressure phase is orthorhombic with $P2_12_12_1$ space group ($a=10.72$ {\AA}, $b=10.78$ {\AA}, $c=10.88$ {\AA}). This is compared to an orthorhombic phase with $Pmc2_1$ space group found beyond 1.3 GPa in the X-ray diffraction experiment \cite{Poswal 2009}.

\begin{figure}[t]
\begin{center}
\includegraphics[width=8.0cm]{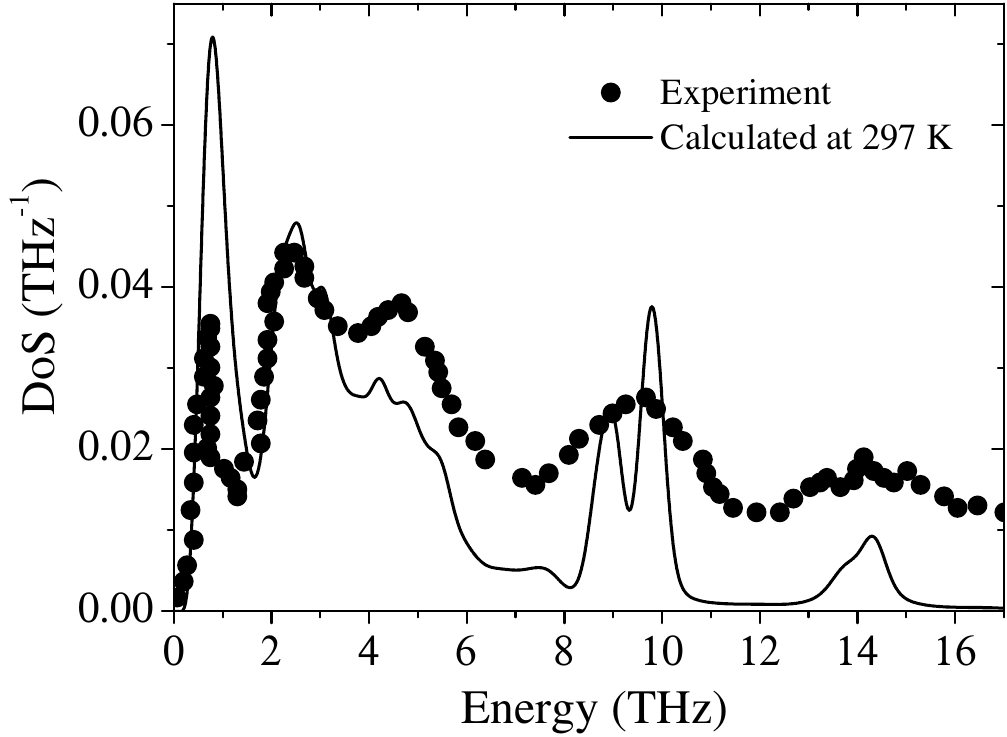}
\end{center}
\caption{\label{fig7} The Zn(CN)$_2$ phonon density of states at 297 K. Solid lines are data from our MD calculations; filled circles are neutron scattering experimental results\cite{Mittal 2011}. The two agree well with one another.}
\end{figure}

\section{Phonons}
\subsection{Density of states}

The Fourier transformation of the atomic velocity auto-correlation function (VACF) gives us the phonon density of states (DoS)~\cite{Martin 1993} of the material. To obtain the DoS at different pressures and temperatures, we used the trajectory data of atoms from MD to calculate the VACF of Zn(CN)$_2$.

The correlation function $C(t)$ at time $t=n \triangle t$ can be expressed as,

\begin{equation}\label{eq3}
C\left( t \right) = \frac{1}{{N(M - n)}}\sum\limits_{j = 1}^N {\sum\limits_{m = 1}^{M - n} {v_j \left( {m\Delta t} \right)} } v_j \left( {m\Delta t + t} \right)
\end{equation}

\noindent where $v_j$ is the velocity component of the $j$th atom. $\Delta t$ is the time interval of $0.02$ ps. $M$ is the total number of time steps. The system was simulated for a total of $50$ ps which corresponds to $M=2500$. The correlation function was calculated for each atom with a time length of $30$ ps ($n=1500$). A Gaussian profile was used before Fourier transformation to suppress the ripple effect caused by time cut-off. The VACF of angular velocities of the C--N rigid rods rotating about their center of mass was also calculated. The Fourier transformation then gave us the DoS of the pure rotational modes of these rigid rods. To obtain the phonon spectrum, the calculated DoS was first multiplied by a weighting factor $4\pi b_k / m_k$ containing the scattering length $b_k$ and the atomic mass $m_k$ of the $k$th atom. Then, in order to mimic experimental resolution \cite{Mittal 2011}, the DoS is convolved with a Gaussian with FWHM of $10$\% of the energy transfer. Fig.~\ref{fig7} shows the good agreement between the calculated spectrum and the experiment \cite{Mittal 2011}.

\begin{figure}[t]
\begin{center}
\includegraphics[width=8.0cm]{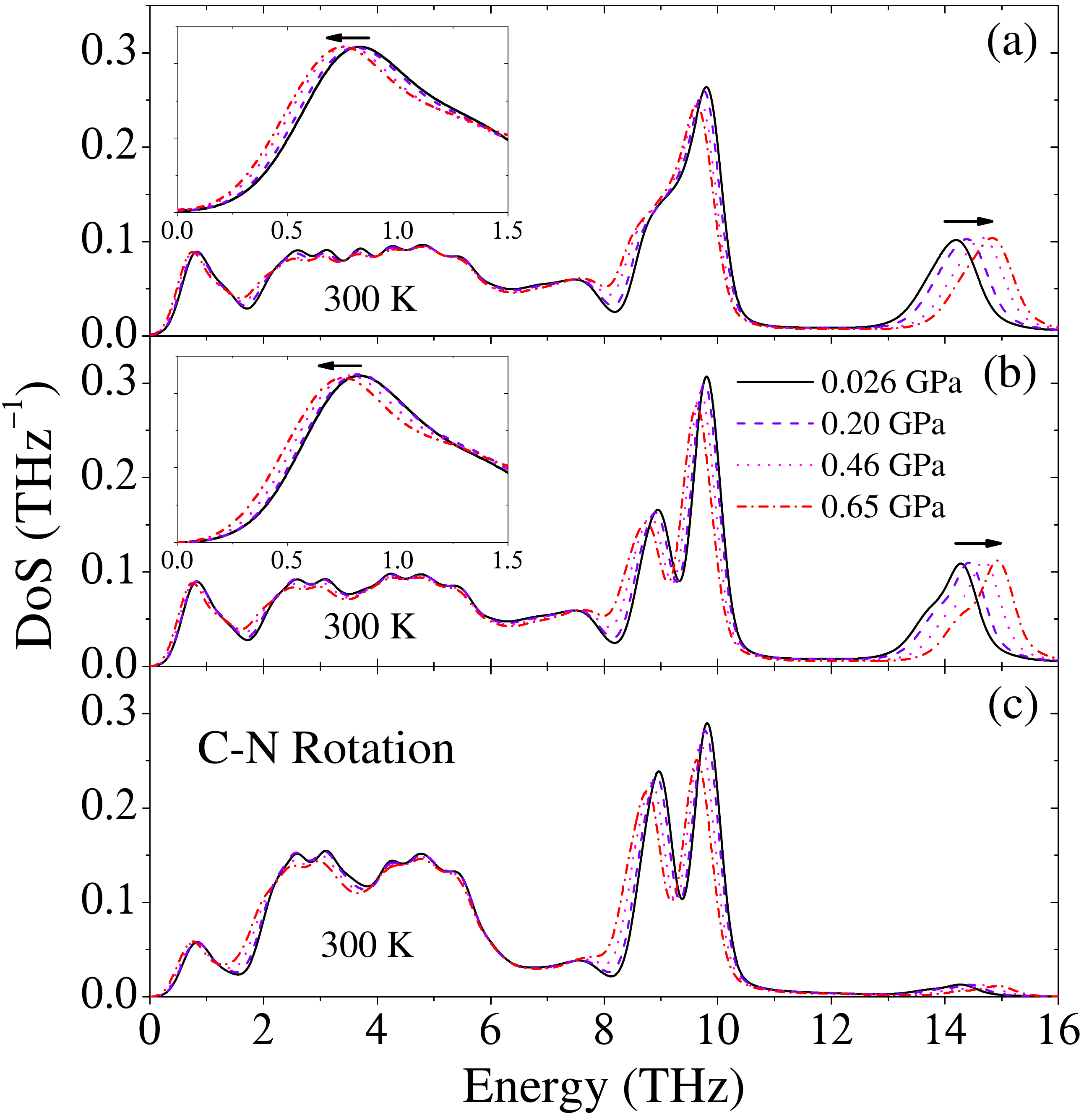}
\end{center}
\caption{\label{fig9} Vibrational DoS calculated using MD at 300 K and at 0.026 GPa, 0.20 GPa, 0.46 GPa and 0.65 GPa. (a) shows the full DoS for the CN-disordered model; (b) shows the full DoS for the CN-ordered model; (c) shows the DoS for only C--N rigid rod rotations in the ordered model. The peaks at around 14 THz, corresponding to pure Zn--C(N) bond flexing, increase in frequency on compression.}
\end{figure}

\begin{figure}[t]
\begin{center}
\includegraphics[width=8.0cm]{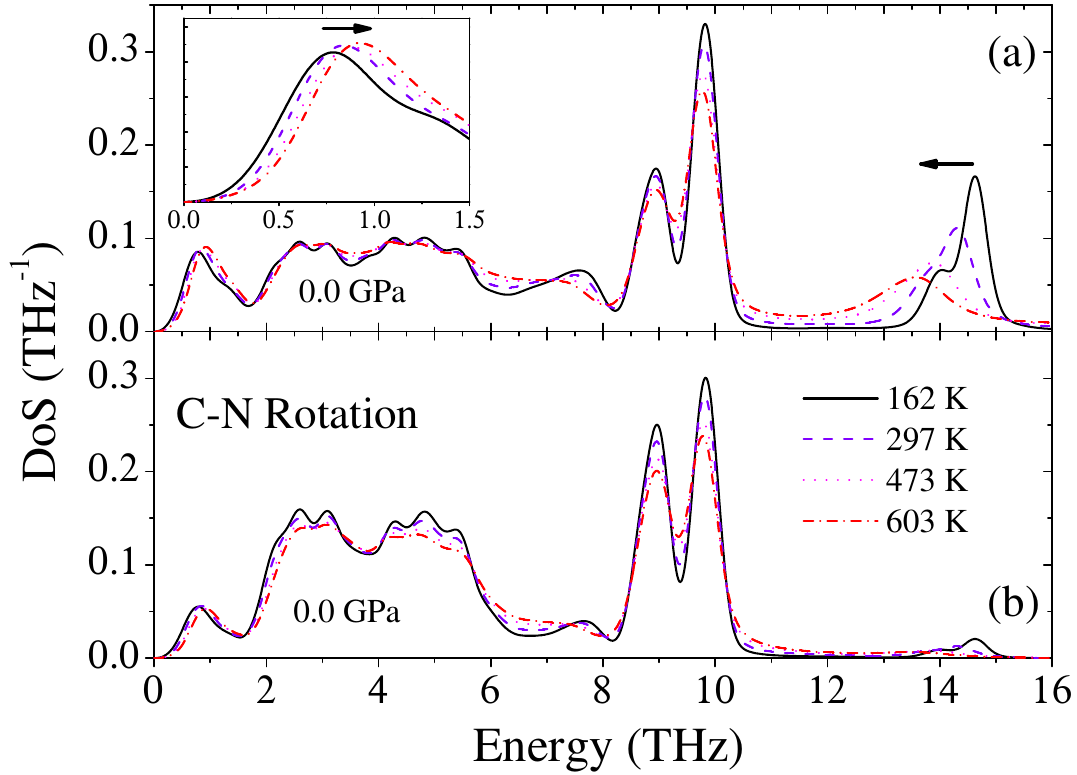}
\end{center}
\caption{\label{fig8} Vibrational DoS calculated using MD at 0.0 GPa and temperatures 162 K, 297 K, 473 K and 603 K. (a) shows the full DoS; (b) shows the DoS for only C--N rigid rod rotations. The peaks in the full DoS at around 14 THz, corresponding to pure Zn--C(N) bond flexing, broaden and decrease in frequency on heating.}
\end{figure}

In order to draw links between the atomic vibrations and the phonon properties, we computed both the overall DoS and the angular DoS of rigid rod C--N, as shown in Fig.~\ref{fig9} and~\ref{fig8}. Fig.~\ref{fig9} (a) and (b) shows the DoS at ambient temperature ($300$ K) under different pressures for the disordered and ordered model, respectively. The acoustic peak around 0.5 THz is softened on compression. The optic peaks around $2.0$ THz and $9.0$ THz follow the same trend, which means that all these modes have negative Gr\"{u}neisen parameters ($\gamma_{\mathbf{k},\lambda}$) and contribute to the NTE of the material. The highest energy peak around $15$ THz is stiffened under compression, suggesting positive $\gamma_{\mathbf{k},\lambda}$ for the Zn--C/N bond flexing modes. The only difference between the disordered model and the ordered one is that the former has broader peaks, especially the merged two peaks around $9.0$ THz (with a broadening of~$\sim 0.8$ THz), due to the `fluffiness' caused by random positions of C and N atoms.

\begin{figure*}[t]
\begin{center}
\includegraphics[width=0.98\textwidth]{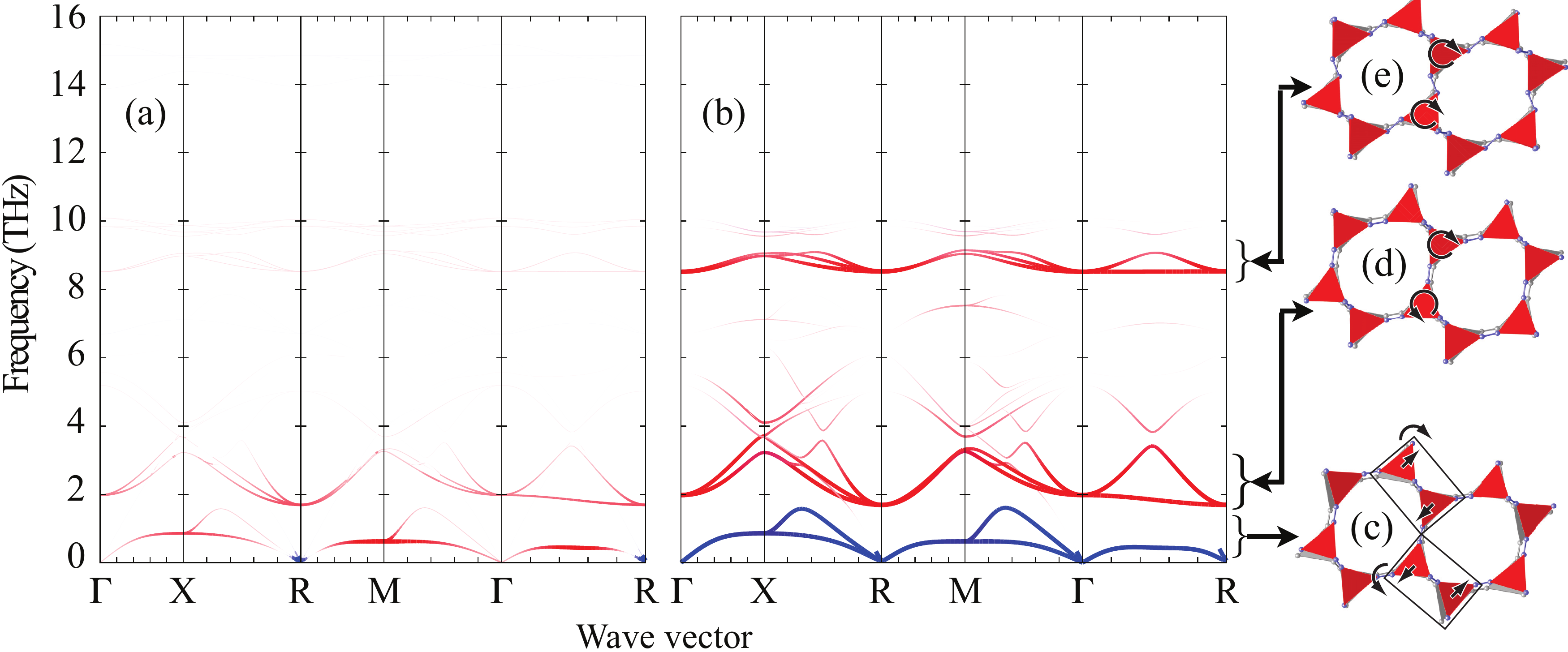}
\end{center}
\caption{\label{fig:dispersioncuves} (a) Dispersion curves coloured according to the corresponding value of the mode Gr\"{u}neisen parameter $\gamma_{\mathbf k,\lambda}$. Red is a negative value of $\gamma_{\mathbf k,\lambda}$ (down to a minimum of $-18$); blue is a positive value of $\gamma_{\mathbf k,\lambda}$ (up to a maximum of 0); white is a $\gamma_{\mathbf k,\lambda}$ value of zero. (b) Dispersion curves coloured according to projection of the eigenvectors onto RUMs. Colour strength corresponds to the degree to which a mode is a RUM; white corresponds to zero RUM character in a given mode. The colour itself corresponds to the nature of the RUM: Red is purely rotational RUM motion; blue is purely translational RUM motion. The corresponding eigenvectors of (c) the translational RUMs (around 0.5 THz), (d) the first rotational RUMs (around 2.0 THz) and (e) the second rotational RUMs (around 9.0 THz), viewed down the [$1,\overline 1,0$] direction with the undistorted structure in grey shown behind. The translational RUMs correspond to lateral translations of the tetrahedra leading to rotations of pairs of tetrahedra. The rotational RUMs correspond to neighbouring tetrahedra rotating in the same or opposite direction.}
\end{figure*}

By comparing Fig.~\ref{fig8}(b) with Fig.~\ref{fig8}(a), we found that about half of the acoustic peak around $0.5$ THz is from vibrations involving rotations of the C--N rod around its centre of mass (bearing in mind that the angular DoS in Fig.~\ref{fig8}(b) is renormalized, so that only the relative heights of the peaks are indicative). Optic peaks round 2.0 and 9.0 THz are hardly changed, because much of their motions can be cast onto the rotations of the C--N rod. The peak with the highest frequency in the overall DoS is from pure Zn--C/N bond flexing --- it completely disappears in the angular DoS. Unlike the acoustic peak, this peak is softened on heating due to the thermal expansion of the Zn--C/N bond, and flattens with elevated temperature due to the finite life time of the corresponding phonon.

\begin{figure}[t]
\begin{center}
\includegraphics[width=8.0cm]{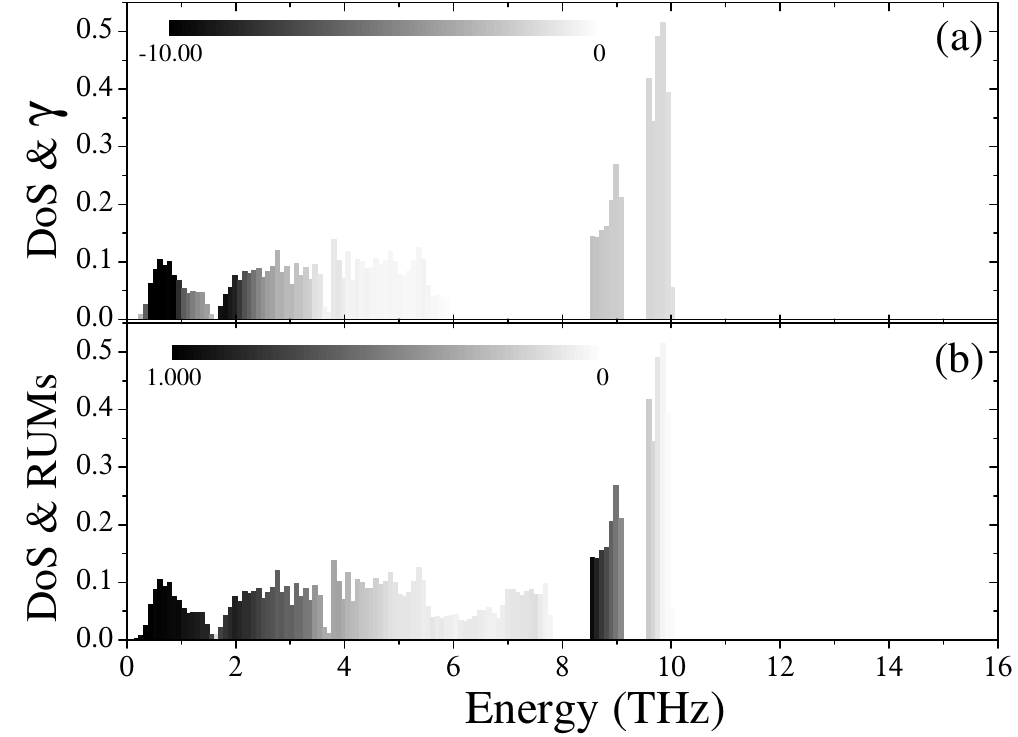}
\end{center}
\caption{\label{fig10} Full DoS calculated using lattice-dynamics. (a) Each bin is coloured according to its average Gr\"{u}neisen parameter: red bins have an average $\gamma_{\mathbf k,\lambda}$ value of $-10$; white bins have a positive average $\gamma_{\mathbf k,\lambda}$. (b) Each bin is coloured according to its average RUM component. Red bins are pure RUMs; blue bins have zero RUM character.}
\end{figure}

\begin{figure}[t]
\begin{center}
\includegraphics[width=8cm]{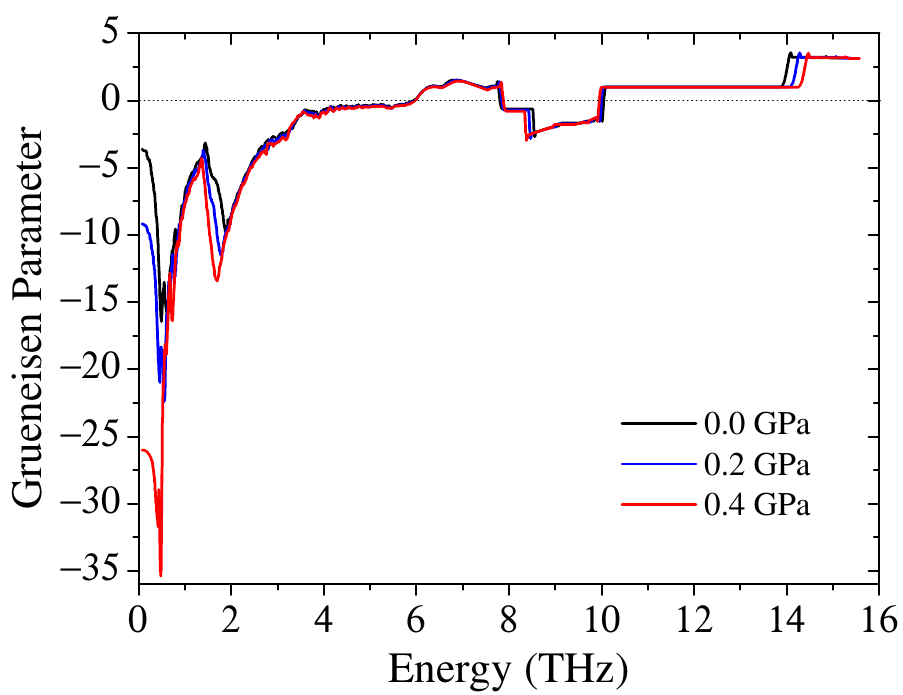}
\end{center}
\caption{\label{fig11} Profiles of mode Gr\"{u}neisen parameters of Zn(CN)$_2$ calculated at three different pressures from lattice dynamics: black curve is 0.0 GPa; blue curve is 0.2 GPa; red curve is 0.4 GPa. The Gr\"{u}neisen parameters of the modes around 0.5 THz, 2.0 THz and 9.0 THz become more negative as elevated pressures, resulting in a more negative overall Gr\"{u}neisen parameter as pressure is increased.}
\end{figure}

\subsection{Rigid unit modes and negative thermal expansion}

Previously, the peak around $0.5$ THz has been found to be the major contributor to the NTE in both experiment \cite{Chapman 2005} and calculations \cite{Zwanziger 2007,Mittal 2011} due to its large negative $\gamma_{\mathbf{k},\lambda}$. However, if one can further identify the corresponding real-space picture of the vibrations, the reason of why the peak has the most negative $\gamma_{\mathbf{k},\lambda}$ compared to the other optic peaks can be revealed.

To understand the nature of various peaks in the DoS, we decided to categorize the vibrational modes in the material using the rigid unit mode model \cite{Giddy 1993}. First, we calculated $\gamma_{\mathbf{k},\lambda}$ from phonon frequencies of expanded and contracted ($\pm$0.01\%) unit-cell volumes. We then coloured the dispersion curves according to both magnitudes and signs of $\gamma_{\mathbf{k},\lambda}$, as shown in Fig.~\ref{fig:dispersioncuves}(a). This representation highlights the most important phonon branches responsible for NTE of the material, namely the low-lying acoustic modes around $0.5$ THz and the lowest-energy optic branches around $2.0$ THz, which both have the most negative $\gamma_{\mathbf{k},\lambda}$ and span the entire Brillouin zone. This was also highlighted in the DoS in Fig.~\ref{fig10}(a), where we coloured the DoS according to the mean values of $\gamma_{\mathbf{k},\lambda}$ for each frequency bin. Then, we calculated the rigid unit modes (RUMs) of Zn(CN)$_2$ using the CRUSH code \cite{Giddy 1993,Hammonds 1994} as the set of eigenvectors of the dynamical matrix whose eigenvalues are zero. We took the dot product between the eigenvectors of the RUMs and the eigenvectors from the lattice-dynamic calculation, and then coloured the dispersion curves in Fig.~\ref{fig:dispersioncuves}(b) by the extent to which each mode eigenvector can be described in terms of correlated whole-body translations (in blue) and rotations (in red) of [Zn(C/N)$_4$] tetrahedra. The DoS in Fig.~\ref{fig10}(b) was also coloured accordingly. One can see that all the modes with negative $\gamma_{\mathbf{k},\lambda}$ that contribute to the NTE of the material are RUMs.

As shown by Fig.~\ref{fig:dispersioncuves}(c), the acoustic modes around $0.5$ THz, like those at M and X, are characterized by translational motions of the rigid tetrahedral units, partly involving angular rotations of the C--N rod. The optic modes around $2.0$ and 9.0~THz can be seen as neighbouring tetrahedral rotating against each other, as respectively shown by Fig.~\ref{fig:dispersioncuves}(d) and (e). The RUM nature of these modes guarantees their low frequencies and large negative $\gamma_{\mathbf{k},\lambda}$. The relatively high frequencies of the optic RUMs is due to the breaking of the Zn--C/N--N/C alignment, and the magnitudes of their negative $\gamma_{\mathbf{k},\lambda}$ suffer accordingly. Study of the eigenvectors also directly revealed that the non-RUM modes around $10$ and $15$ THz correspond to the pure angular vibration of C/N--Zn--N/C within the tetrahedra and the pure bond flexing of Zn--C/N, respectively.

We also found that, besides the negative $\bar{\gamma}$, Zn(CN)$_2$ also has negative $\partial \bar{\gamma} / \partial P$ and $\partial^2 \bar{\gamma} / \partial P^2$. Fig.~\ref{fig11} shows the profiles of $\gamma_{\mathbf{k},\lambda}$ up to $0.4$ GPa calculated in HLD. It is clear that $\bar{\gamma}$ becomes more negative on compression due the contributions from the RUMs around $0.5$ THz, $2.0$ THz and $9.0$ THz, with the translational RUMs having the most negative $\gamma_{\mathbf{k},\lambda}$ (refer to Fig.~\ref{fig10}(a)) contributing the most. The negativity of the pressure derivatives of $\bar{\gamma}$ would result in negative $B^\prime_0$ at non-zero temperatures \cite{Fang zeolite}.

\begin{figure}[t]
\begin{center}
\includegraphics[width=8.0cm]{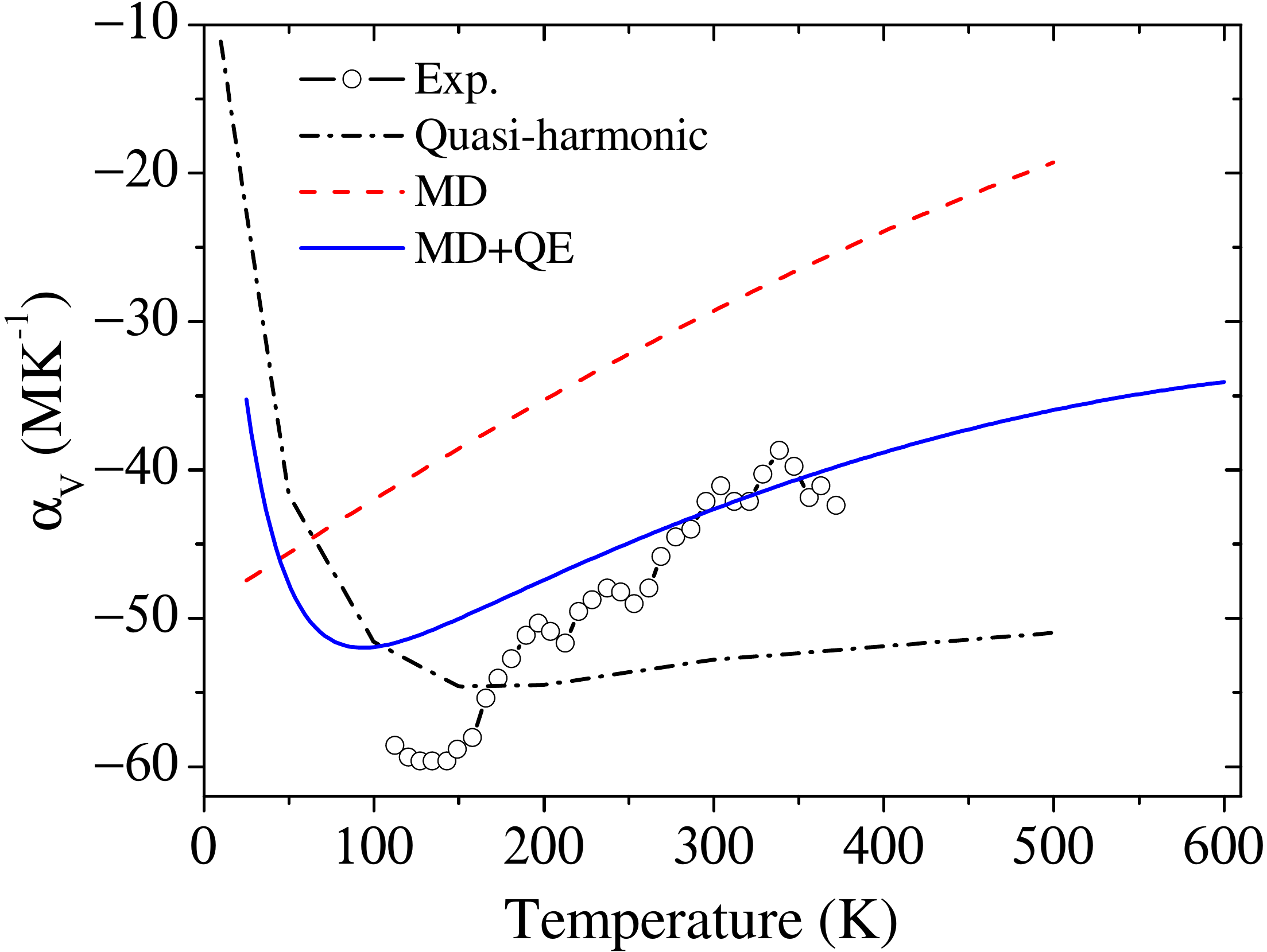}
\end{center}
\caption{\label{figA1} Calculated temperature dependence of the NTE with both anharmonicity and quantum effects included (`MD+QE', solid curve), together with the results from MD (dashed curve) and QHLD (dash-dot curve). The former clearly has a better agreement with the experiment (empty circle)~\cite{Chapman 2005}.}
\end{figure}

\subsection{Anharmonicity and quantum effects}

We found that the acoustic peak in the DoS (see Fig.~\ref{fig10}) with frequencies less than 1 THz ($\sim 50$ K) accounts for half of the NTE ($\alpha=-52.8$ MK$^{-1}$ is reduced to $\alpha=-26.8$ MK$^{-1}$ when excluding these modes, calculated by Eq.~\ref{eq1} at 300 K in QHLD). This suggests that even at low temperatures these modes will not be `frozen' out and can still be excited and contribute to NTE and its relevant properties such as pressure-enhanced NTE and pressure-induced softening of the material. Thus, the classical MD results at low temperatures would not have too much difference from the real quantum picture and can give a good qualitative agreement with experiments.

\begin{figure}[t]
\begin{center}
\includegraphics[width=8.0cm]{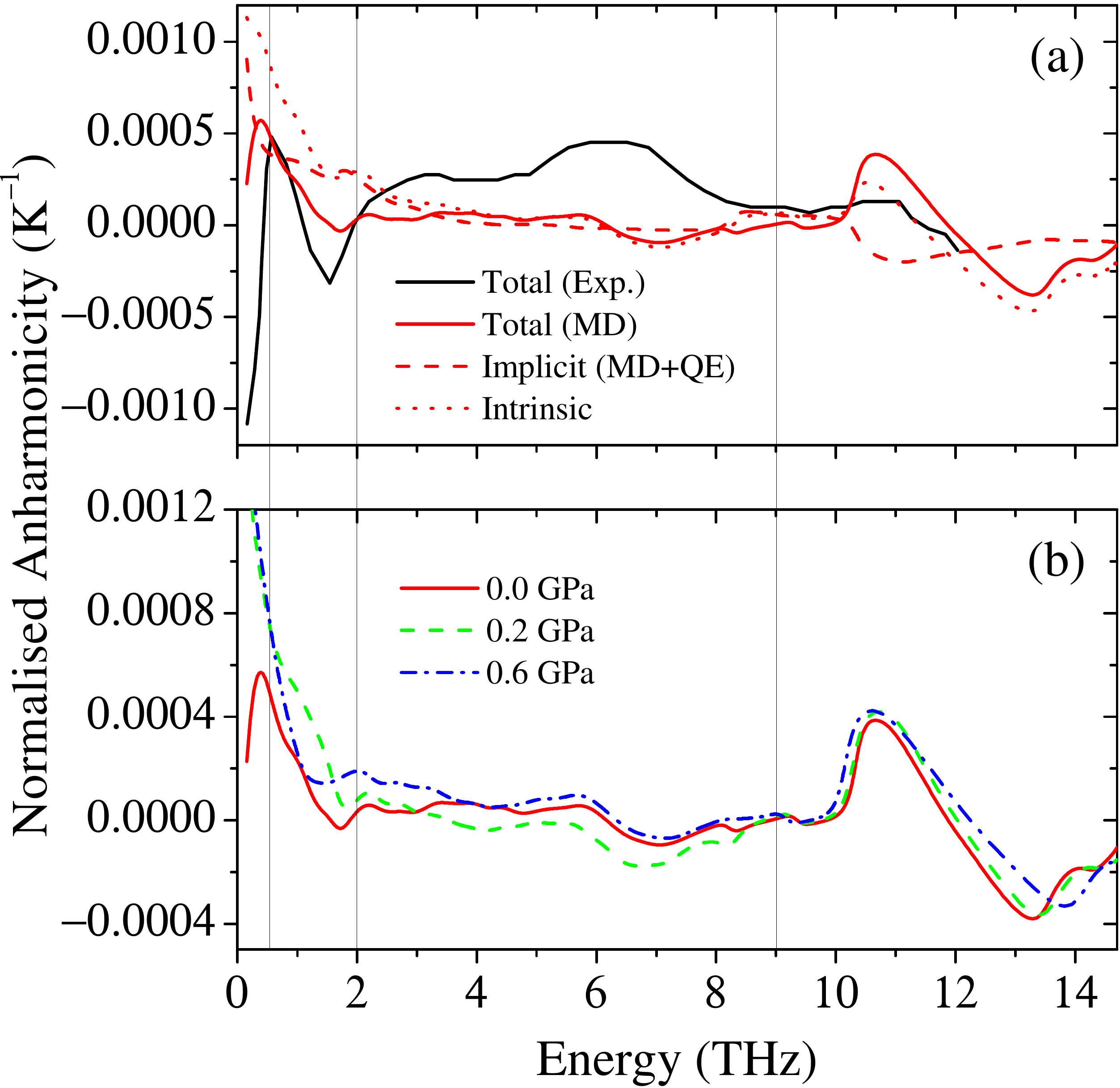}
\end{center}
\caption{\label{figA2} (a) Calculated normalized (in red solid), implicit (in red dash) and intrinsic (in red dots) anharmonicity using cumulative distributions of DoS from the MD at 180 and 240 K, compared to the total anharmonicity from the experiment DoS at the same two temperatures~\cite{Mittal 2011}. (b) Calculated normalized anharmonicity (using cumulative distributions of DoS from MD at 180 and 240 K) at different pressures. The important modes around 0.5, 2.0 and 9.0 THz have positive anharmonicity with the translational RUMs (around 0.5 THz) shows the largest value compared to others. The modes around 15 THz corresponding to Zn--C(N) bond flexing show negative anharmonicity. The anharmonicity of all these modes is enhanced by compression.}
\end{figure}

However, with the following method, we can include the effect of both anharmonicity and quantum effects in the temperature dependence of NTE of the material. First, at a certain temperature, we calculate two DoS from MD for two adjacent volumes (with $0.5 \%$ difference). Then we can obtain phonon frequencies and mode Gr\"{u}neisen parameters by using the cumulative distributions of these two DoS. Finally, $\alpha_\mathrm{V}$ can be calculated by Eq.~\ref{eq1}. We then repeat this process at different temperatures up to 600 K, and obtain the temperature dependence of NTE of the material.

As such, anharmonicity is accounted for by the use of the DoS from the MD, while quantum effects are included in the formalism of Eq.~\ref{eq1} to calculate $\alpha_\mathrm{V}$. Together with the direct MD and QHLD results, the temperature dependence of the NTE from this method (`MD+QE') is shown in Fig.~\ref{figA1}. At low temperatures, the curve acts like the QHLD result due to quantum quenching. At high temperatures, the curve becomes less negative like MD due to anharmonicity, making the curve in a better agreement with the experiment.

From the same analysis, we can also obtain the temperature and pressure dependence of anharmonicity. The normalized anharmonicity \cite{Ravindran 2003} measuring the change of mode frequency with temperature at constant pressure is defined as

\begin{eqnarray}\label{eq7}
\left. {\frac{1}{{\omega _s }}\frac{{\partial \omega _s }}{{\partial T}}} \right|_p  = \left. {\frac{1}{{\omega _s }}\frac{{\partial \omega _s }}{{\partial T}}} \right|_V  - \gamma _s \alpha _\mathrm{V},
\end{eqnarray}

\noindent where, on the right-hand side, the first term is the intrinsic anharmonicity and the second term is the contribution from the contraction of lattice (implicit). By using the DoS from the MD at 180 and 240 K, the normalized anharmonicity of each mode was calculated, as shown in Fig.~\ref{figA2}(a). The results of the important modes around 0.5, 2.0 and 9.0 THz, as indicated by the vertical lines, agree quite well with the experiment (using DoS at the 180 and 240 K)~\cite{Mittal 2011}. The implicit anharmonicity was calculated using the mode Gr\"{u}neisen parameters and the `MD+QE' value of $\alpha_\mathrm{V}$ at 180 K. The intrinsic anharmonicity was then obtained from Eq.~\ref{eq7}.

\begin{figure}[t]
\begin{center}
\includegraphics[width=8.0cm]{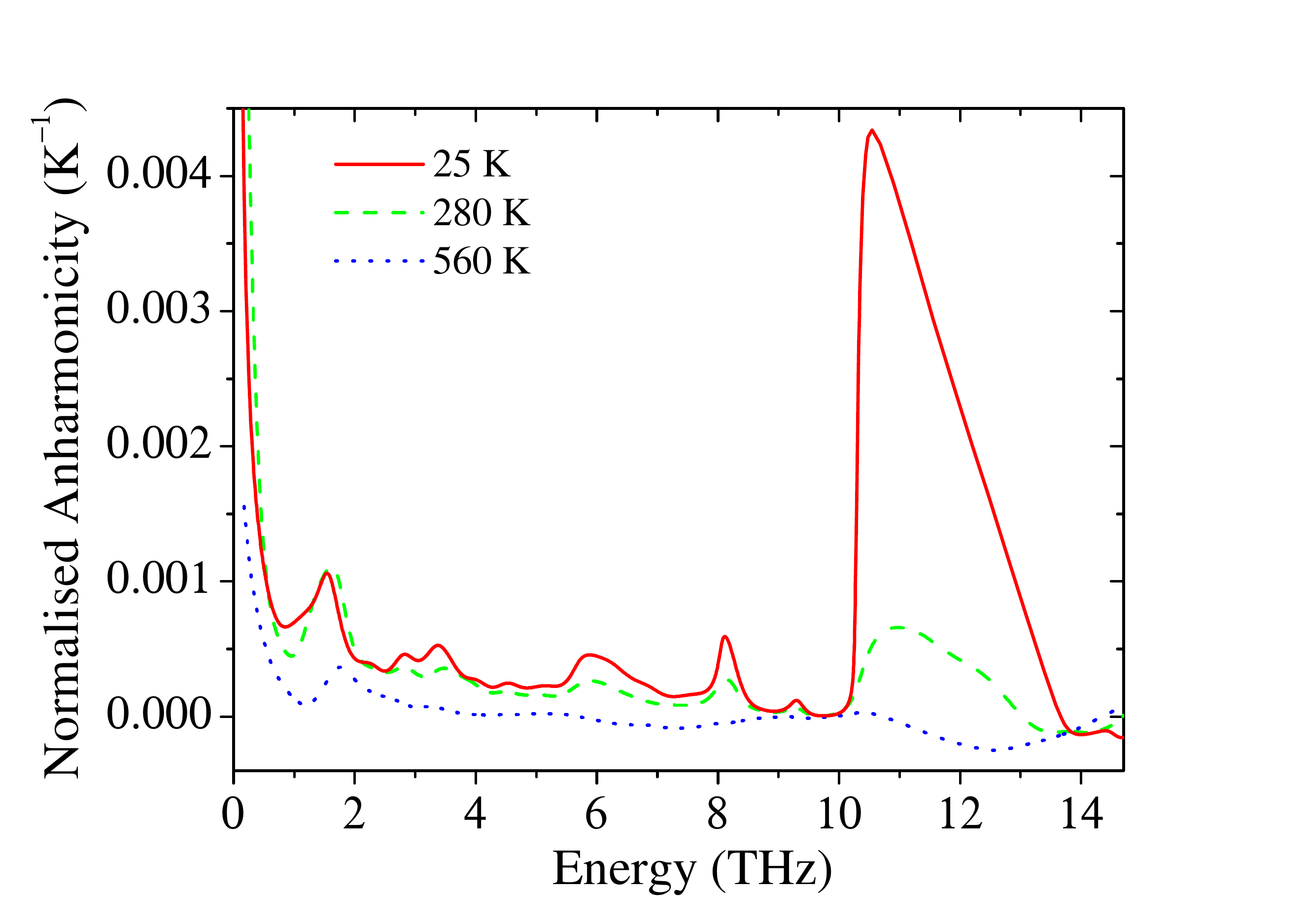}
\end{center}
\caption{\label{figA3} Temperature dependence of the normalized anharmonicity at 0.0 GPa. At 25, 280 and 560 K, calculations were conducted using the cumulative distributions of couples of DoS at 25/45 K, 280/300 K and 560/590 K, respectively. The general trend is that the mode frequencies change less rapidly with temperature on heating.}
\end{figure}

The normalized anharmonicity at different pressures is shown in Fig.~\ref{figA2}(b). As mentioned in the former sections, the modes around 0.5, 2.0 and 9.0 THz, corresponding to the translational and rotational RUMs, respectively, are stiffened on heating with positive normalized anharmonicity. Among these, the translational RUMs (around 0.5 THz) show the largest normalized anharmonicity of more than $4\times10^{-4}$ K$^{-1}$ at 0.0 GPa. The modes around 15 THz corresponding to pure bond flexing of Zn--C(N) are softened on heating with negative anharmonicity. The normalized anharmonicity of all these peaks is strengthened on compression.

We further calculated the normalized anharmonicity at low (25 K), medium (280 K) and high (560 K) temperatures to see its temperature dependence, as shown in Fig.~\ref{figA3}. At each temperature, two DoS with temperature difference less than 30 K are used. The low-temperature (25 K) value of the anharmonicity of the modes around 0.5 THz is $1.0\times10^{-3}$ K$^{-1}$, in good agreement with the experimental value of $1.1\times10^{-3}$ K$^{-1}$ in Ref.~\onlinecite{Chapman 2006}. The figure also shows that the mode frequency would change less rapidly on heating at high temperatures. The same trend is seen for the 0.5 THz peak in the experiment in Ref.~\onlinecite{Chapman 2006}.

\section{Structural study}
\subsection{The local picture on compression and heating}

\begin{figure}[t]
\begin{center}
\includegraphics[width=8.0cm]{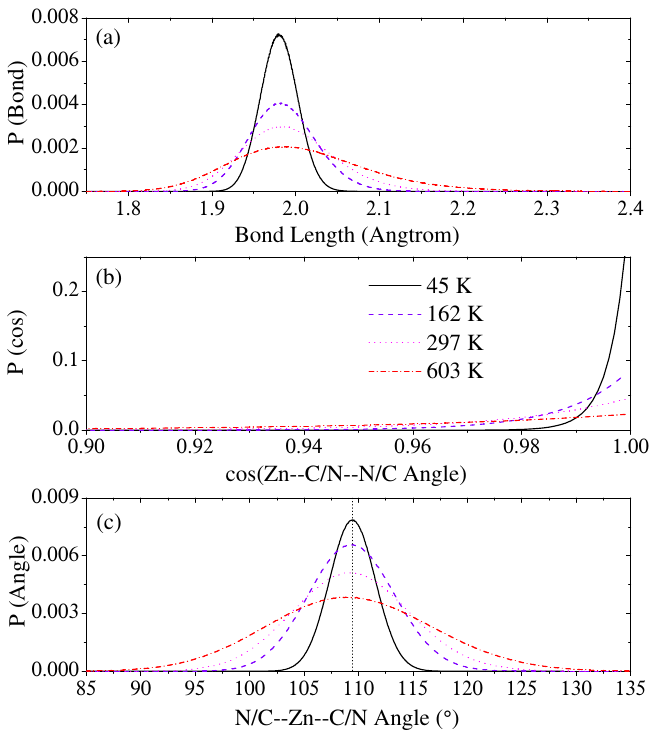}
\end{center}
\caption{\label{fig12} Distributions of (a) the Zn--C/N bond length; (b) the cosine of Zn--C/N--N/C angle distortion; (c) the N/C--Zn--C/N angle within the tetrahedral unit at 0.0 GPa.}
\end{figure}

In real space, variations of geometrical features of the material, such as the Zn--C/N bond length, the N/C--Zn--C/N angle within the tetrahedral unit, and the Zn--C/N--N/C angle are important for us to build a local picture of the system subject to both compression and heating.

Distributions of these quantities, as shown in Fig.~\ref{fig12}, were obtained from the atomic trajectory data of the MD. The large spread of the distributions suggests large vibrations of these quantities at high temperature. The slightly expansion-biased broadening of the bond-length distribution on heating indicates an enhanced thermal expansion in the bond. We found that the deviation of the average N/C--Zn--C/N angle in the tetrahedral unit from its equilibrium of $109.47^\circ$ is trivially small ($\sim 0.2^\circ$) even at very high temperature ($\sim 600$ K) and pressure ($\sim 0.6$ GPa).

\begin{figure}[t]
\begin{center}
\includegraphics[width=8.0cm]{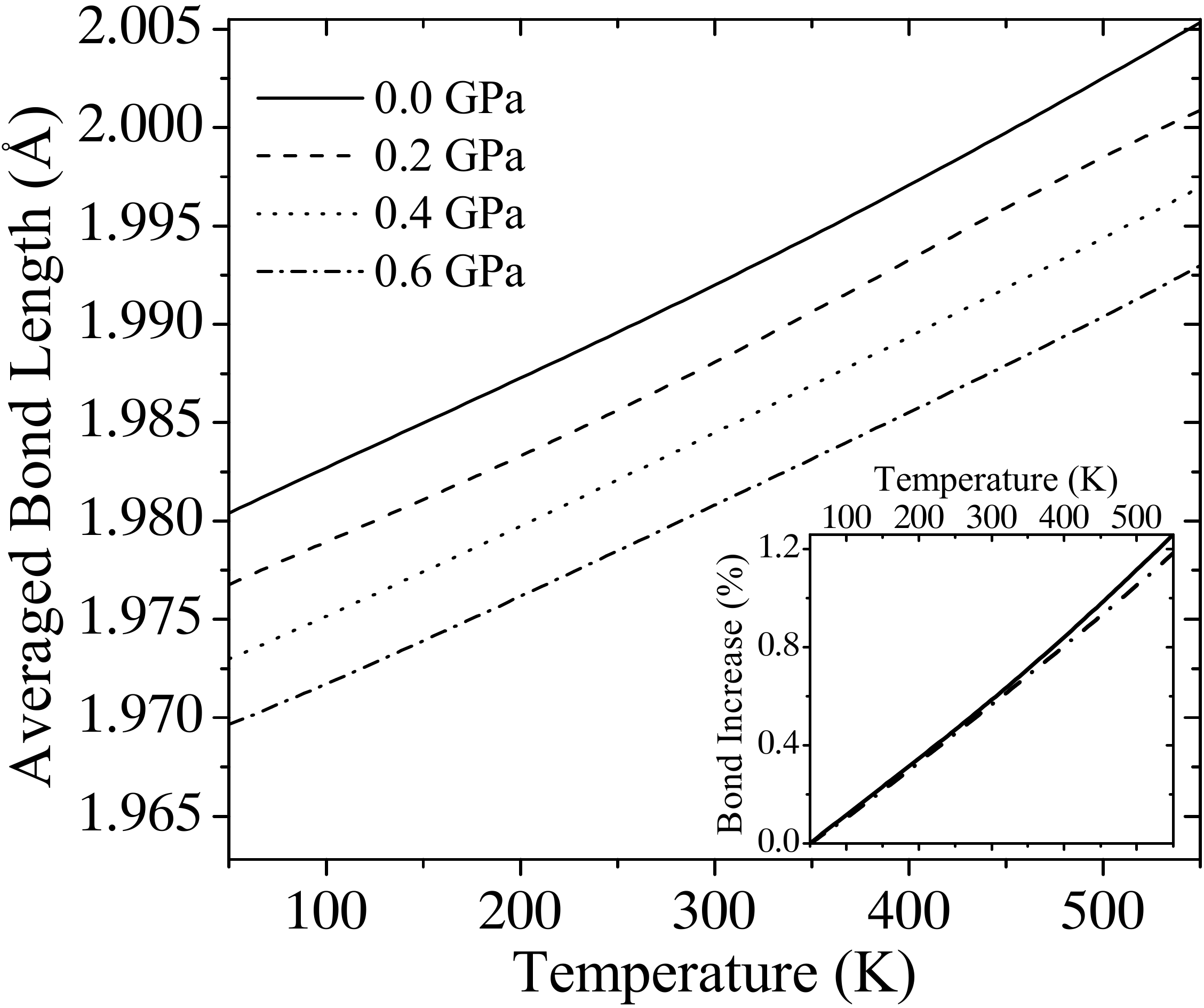}
\end{center}
\caption{\label{fig13} Averaged Zn--C/N bond length as a function of temperature, calculated from MD at pressures $0.0$~GPa, $0.2$~GPa, $0.4$~GPa and $0.6$~GPa. The bond length increases with elevated temperature. The inset shows superlinear behaviour of the bond length on heating.}
\end{figure}

\begin{figure}[t]
\begin{center}
\includegraphics[width=7.6cm]{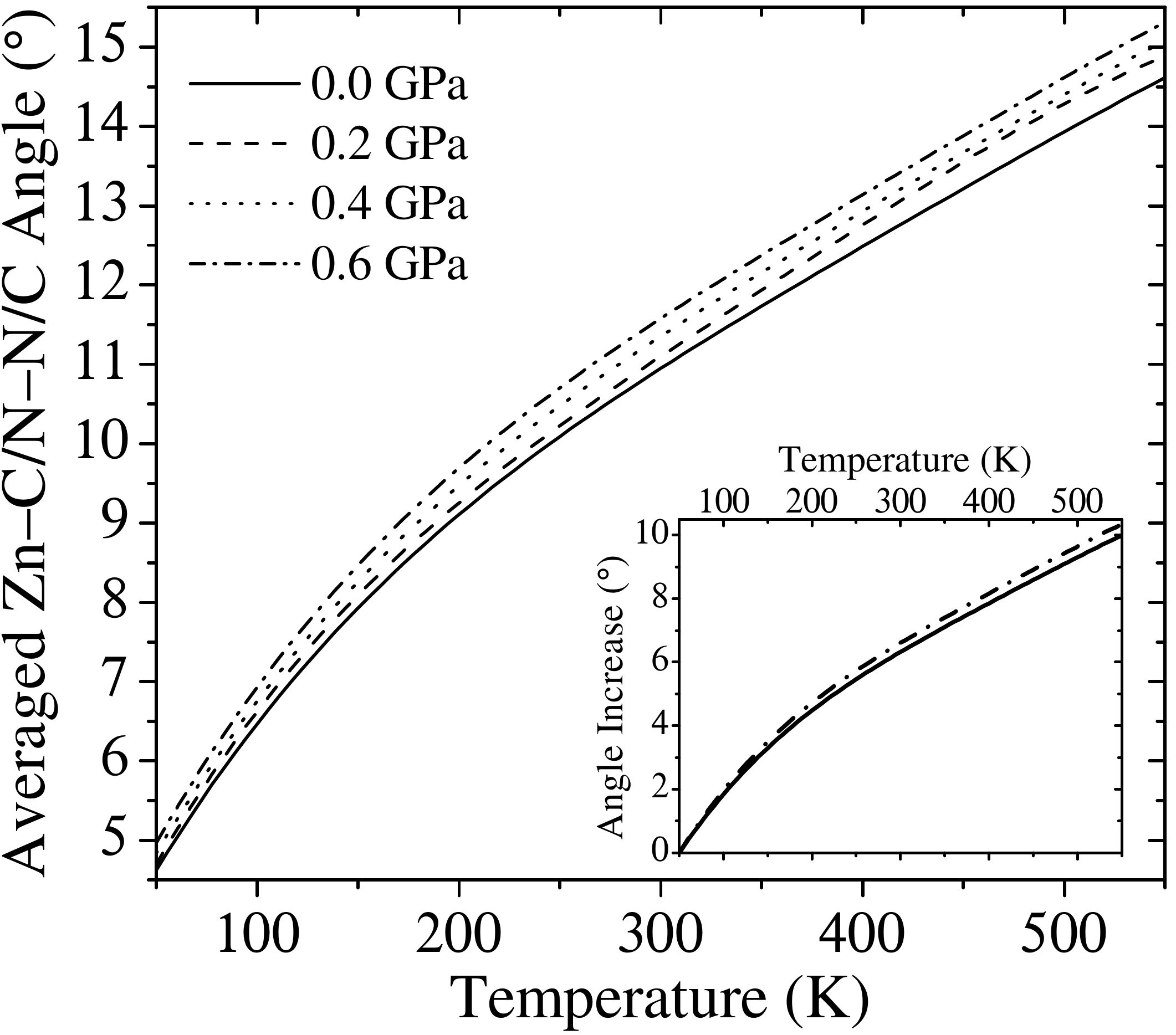}
\end{center}
\caption{\label{fig13-2} Averaged Zn--C/N--N/C angle as a function of temperature, calculated from MD at pressures $0.0$~GPa, $0.2$~GPa, $0.4$~GPa and $0.6$~GPa. The angle distortion increases with elevated temperature and pressure. The inset shows the sublinear behaviour of the angle distortion on heating.}
\end{figure}

The average bond length and the average angle distortion in Zn(CN)$_2$ as functions of both temperature and pressure are shown in Fig.~\ref{fig13} and~\ref{fig13-2}, respectively. Compression progressively increases the Zn--C/N--N/C angle with elevated temperature. At very low and zero temperature, the trend is that the angle will be hardly changed by pressure. This is exactly what expected in the previous section where we suggested that, at zero temperature, the volume change of the material due to the pressure arises solely from the compression of the Zn--C/N bonds, resulting in positive $B_0^{\prime}$. It is when the Zn--C/N--N/C angle starts to increase under compression and contribute to the volume change that the material shows negative $B_0^{\prime}$, i.e. the pressure-induced softening. The inset of Fig.~\ref{fig13} shows the increase of bond length with temperature. The superlinear behaviour indicates the softening of the bond at higher temperature due to the thermal expansion. In Fig.~\ref{fig13-2}, the inset shows the increase of the Zn--C/N--N/C angle with elevated temperature. The sublinear behaviour suggests that the angle will become more rigid, which should result in a less negative $B_0^{\prime}$ at high temperatures \cite{our_Exp}. Both the superlinearity and the sublinearity in the plots rely on the capture of the anharmonicity of the material.

\begin{figure}[t]
\begin{center}
\includegraphics[width=8cm]{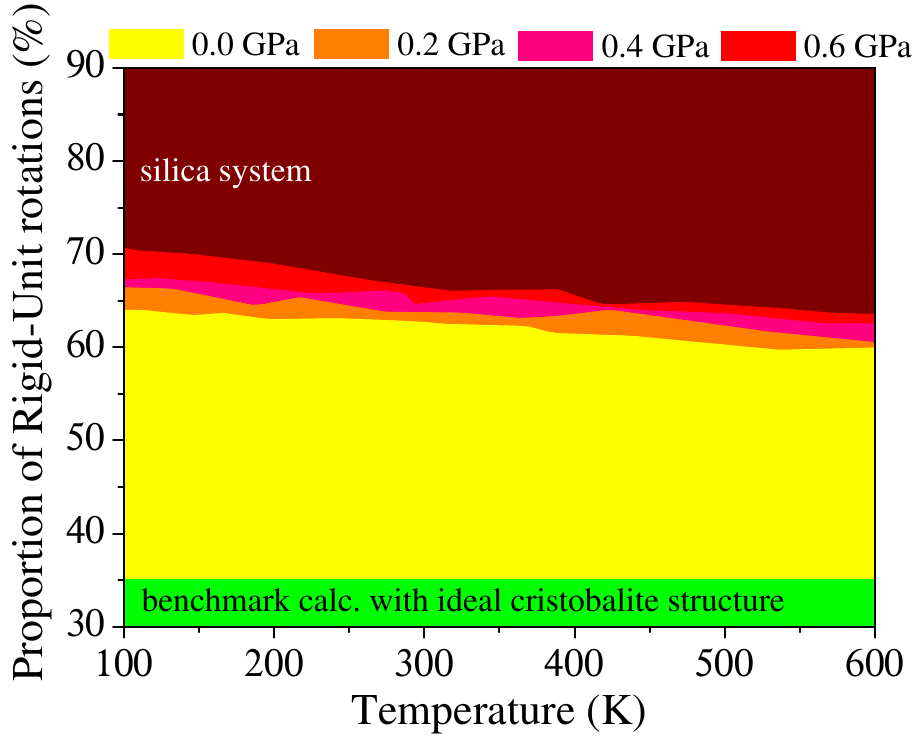}
\end{center}
\caption{\label{fig14} Plot showing the proportion of rotational RUMs present in Zn(CN)$_2$ as a function of temperature, calculated using GASP and data from MD simulations. Data is presented for four pressures plus two additional benchmark systems. The green region is the rotational component present in the ideal cristobalite structure; the yellow region is the additional rotational component present in Zn(CN)$_2$ at 0.0 GPa; the orange region is the additional component present at 0.2 GPa, the pink region is the additional component at 0.4 GPa; the red region is the additional component at 0.6 GPa and the dark red region is the additional component present in amorphous silica. The results clearly show the enhancement of the rotational RUMs in Zn(CN)$_2$ under pressure, as well as the reduction of the rotational RUMs on heating.}
\end{figure}

To see the real-space picture of RUMs in Zn(CN)$_2$, we quantified the proportion of thermal-excited rigid-unit rotations at different pressures. Using our GASP code \cite{Wells 2002a, Wells 2002b, Wells 2004}, we compared ten snapshots of an MD simulation, where each snapshot is separated by 2 ps, with the ideal structure. This is repeated for various pressures and temperatures. GASP, using geometric algebra, can partition the atomic displacements for every comparison made into the mean squared rigid-tetrahedron rotations, translational displacements and unit deformations. We then computed the average proportion of rigid-unit rotations at each temperature. To exclude those rigid-unit rotations due to pure topology reasons, i.e. rotations that accidentally maintain the shape of the tetrahedral unit under thermal excitation but are not because of the features of motion, we set up a benchmark calculation using an ideal cristobalite structure without any interactions other than bonds to hold Zn--C/N and C--N. The reason to use the single-framework lattice is to avoid the problem of two interpenetrating frameworks crushing into each other in the MD due to the lack of long-distance interactions. Fig.~\ref{fig14} shows the results at different pressures and temperatures with coloured areas. The proportion of the rigid-unit rotation of a real silica system is given in the plot as a comparison. The figure suggests that compression will enhance the rigid-unit rotations. At ambient temperature $\sim 300$ K, for example, the average proportions are $63 \%$ at $0.0$ GPa, $64 \%$ at $0.2$ GPa, $65 \%$ at $0.4$ GPa and $66 \%$ at $0.6$ GPa, compared to $35 \%$ of the benchmark and $90 \%$ of the silica system. Another important point is that, at certain pressure, the proportion of rigid-unit rotation will decrease with elevated temperature. This trend corresponds to the peaks around $0.5$, $2.0$ and $9.0$ THz in DoS stiffened on heating, as shown in Fig.~\ref{fig8}. The Gr\"{u}neisen parameters of these modes will consequently become less negative, and so does the coefficient of thermal expansion.

\subsection{The nearest-neighbour Zn$\ldots$Zn distance}

There is a puzzling observation from the X-ray pair distribution function measurements of Zn(CN)$_2$ \cite{Chapman 2005} that instantaneous Zn$\ldots$Zn distances contract less rapidly on heating than does the cell length, while crystallographically the two should be linked. The explanation relies on the nature of the acoustic modes as translational RUMs.

The acoustic modes around $0.5$ THz are translational RUMs which count for half of the NTE of the material as mentioned in Sec. VII. Unlike rotational RUMs around $2.0$ and $9.0$ THz that will reduce the nearest Zn$\ldots$Zn distance, the translational RUMs correspond to collective translations of the neighbouring rigid units that moves zinc atoms off site and retains the distance of the nearest-neighbour zincs, as seen in Fig.~\ref{fig:dispersioncuves}(c). This kind of vibration involves the rotation of C--N rod around its centre of mass, consistent with the previous finding in the DoS that part of the acoustic modes are from the C--N rod rotations.

\begin{figure}[t]
\begin{center}
\includegraphics[width=8cm]{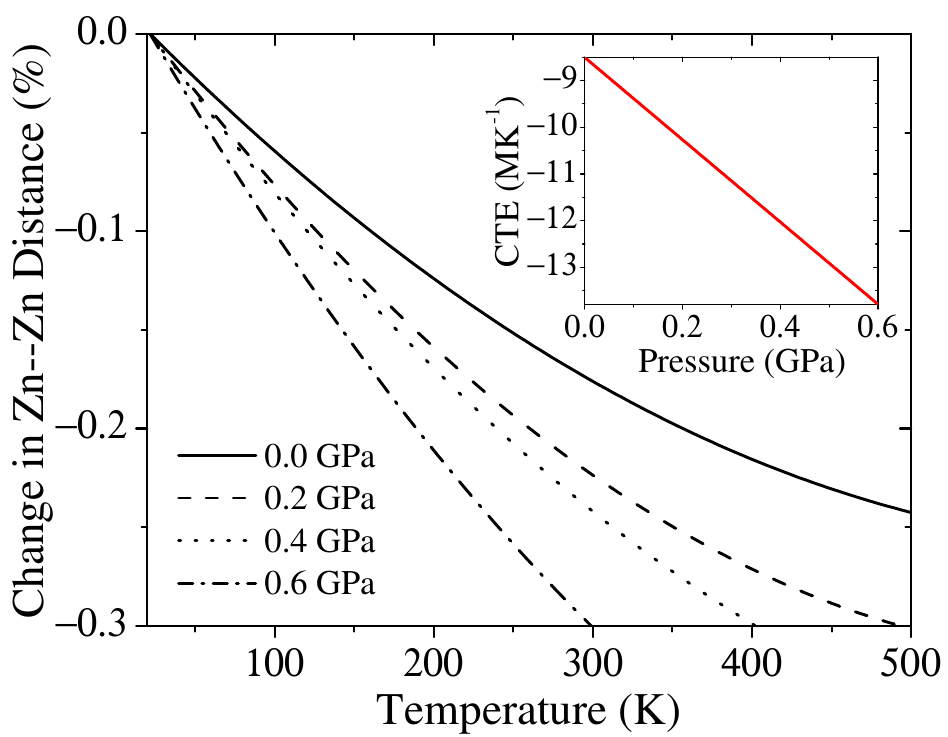}
\end{center}
\caption{\label{fig15} Variation of the averaged nearest-neighbour Zn$\ldots$Zn distance, calculated using MD. Distances have been calculated for the temperature range of 0--500 K and at pressures 0.0 GPa, 0.2 GPa, 0.4 GPa and 0.6 GPa. The inset shows the CTE of the nearest-neighbour Zn$\ldots$Zn distance as a function of pressure.}
\end{figure}

We calculated the average distance of the nearest Zn$\ldots$Zn using the the trajectory data from the MD simulations. Fig.~\ref{fig15} shows the temperature dependence of the average distance under different pressures. At zero pressure, the ratio between the linear coefficient of thermal expansion (CTE) of the averaged Zn$\ldots$Zn distance, $\alpha_{\mathrm{ZnZn}}$, and the overall linear CTE of the material, $\alpha$, is 0.67, with $\alpha_{\mathrm{ZnZn}}=-8.5$\,MK$^{-1}$ and $\alpha=-12.62$\,MK$^{-1}$. This result agrees well with the experimental ratio \cite{Chapman 2005} of 0.71, with $\alpha_{\mathrm{ZnZn}}= -12.42(12)$\,MK$^{-1}$ and $\alpha = -17.40(18)$\,MK$^{-1}$.

\section{DISCUSSIONS AND CONCLUSIONS}

This study shows that almost all the modes responsible for the NTE of Zn(CN)$_2$ are RUMs. We managed to categorize these modes in terms of their vibrational motions. The TA modes around 0.5 THz spanning to the lowest energy correspond to collective motions of Zn--CN--Zn as a rigid body, which can keep the distance of the nearest-neighbour zincs. The low energy hence the most negative Gr\"{u}neisen parameters of these modes are due to this kind of collective motion without involving relatively high-energetic angle bending in the Zn--CN--Zn linkage. These modes contribute half of the NTE of the material. The optic modes around 2.0 THz and 9.0 THz correspond to rotations of the neighbouring tetrahedral units against each other involving angle bending in the Zn--CN--Zn linkage, resulting in higher mode energy and less negative Gr\"{u}neisen parameters.

Although the increase of pressure or temperature would both result in volume contraction in Zn(CN)$_2$, the pressure and temperature dependence of the NTE in Zn(CN)$_2$ are totally different. Increasing temperature stiffens the low-frequency peaks and softens the high-frequency peaks in the DoS (Fig.~\ref{fig8} and Fig.~\ref{figA2}) accompanied by the reduction of NTE, while compression would soften the low-frequency peaks and stiffen the high-frequency peaks (Fig.~\ref{fig9}) resulting in NTE enhancement. Raising temperature slows the mode softening caused by compression, and postpones the phase transition. The enormous decrease of the bulk modulus on heating contrasts the small change of the bulk modulus on compression, which disobeys Birch's law of corresponding states.

The pressure and temperature dependence of the Zn--C/N bond and the N/C--Zn--C/N angle are also intriguing. The large vibrational amplitude of the N/C--Zn--C/N angle extends the Zn--C/N bond, which can be seen (under constant pressure) in Fig.~\ref{fig13}. The superlinearity suggests an enhancement of the thermal expansion in the bond with more distorted angle at higher temperature. The sublinearity of the average angle distortion corresponds to the stiffened RUMs involving the rotational vibrations of the C--N rods.

The ability to carry out MD for Zn(CN)$_2$ using the potential model is vital in this study. It allows us to capture the anharmonicity to reproduce the exotic properties of the material, and to study their pressure and temperature dependence. The origins of various properties are revealed by linking features in both energy and real space.

One example is the pressure-enhanced NTE, which has been well reproduced in Fig.~\ref{fig3}(a). This behaviour is a natural result followed by softening of the bulk modulus on heating (Eq.~\ref{eq4}), and is linked to the feature in energy space that the modes around $0.5$, $2.0$ and $9.0$ THz are softened under compression (as in Fig.~\ref{fig9}), as well as the rising proportion of the rigid-unit rotations in real space on compression (as in Fig.~\ref{fig14}).

Another example is the reduction of NTE with elevated temperature. Stiffening of the modes around $0.5$, $2.0$ and $9.0$ THz on heating (as shown in the DoS in Fig.~\ref{fig8}) makes their Gr\"{u}neisen parameters less negative, hence the reduction of NTE in the material. In real space, Fig.~\ref{fig14} clearly shows the trend of reduction in the proportion of rigid-unit rotations on heating.

The third example is the temperature dependence of the bulk modulus and its first derivative $B_0^{\prime}$. The large decrease of the bulk modulus on heating is due to the involvement of the Zn--C/N--N/C angular vibrations at non-zero temperature and the softening of the Zn--C/N bond, which can be seen both in the DoS (Fig.~\ref{fig8}(a)) and in the real-space picture of Fig.~\ref{fig12}, ~\ref{fig13} and~\ref{fig13-2}. As shown in Fig.~\ref{fig13-2}, the average Zn--C/N--N/C angle starts to possess an almost linear pressure dependence at medium temperature and contributes to the pressure-induced volume change of the material. According to a simple geometrical relation \cite{Heine 1999}, the relative decrease in cell is roughly proportional to the Zn--C/N--N/C angle squared, hence to the pressure squared, resulting in negative $B_0^{\prime}$. However, this contribution from the pressure-induced change in the Zn--C/N--N/C angle to the volume contraction will be hindered at high temperature due to the anharmonic stiffening of the corresponding modes (Fig.~\ref{fig8}(a) and Fig.~\ref{figA2}). On the other hand, the contribution from Zn--C/N bond compression will increase due to the thermal softening of the bond. The combined effect of these two trends is expected to result in a less negative $B_0^{\prime}$ at high temperatures \cite{our_Exp}.

\begin{acknowledgements}
We gratefully acknowledge financial support from the Cambridge International Scholarship Scheme (CISS) of the Cambridge Overseas Trust and Fitzwilliam College of Cambridge University (HF), NERC and CrystalMaker Software Ltd. (LHNR). The MD simulations were performed using the CamGrid high-throughput environment of the University of Cambridge. The interatomic potential was developed through our membership of the UK HPC Materials Chemistry Consortium, funding by EPSRC (EP/F067496), using the HECToR national high-performance computing service provided by UoE HPCx Ltd at the University of Edinburgh, Cray Inc and NAG Ltd, and funded by the Office of Science and Technology through EPSRC's High End Computing programme.
\end{acknowledgements}



\end{document}